%% file: main.tex
\PassOptionsToPackage{table,xcdraw}{xcolor}
\documentclass[manuscript, screen, nonacm]{acmart}

\AtBeginDocument{%
  }

\usepackage{algorithm}
\usepackage{algpseudocode}
\usepackage{tikz}
\usepackage{booktabs}
\usepackage{graphicx}
\usepackage{subcaption}
\usepackage{lscape}
\usepackage{multicol}
\usepackage{multirow}

\newcommand*\circled[1]{\tikz[baseline=(char.base)]{
            \node[shape=circle,draw,inner sep=2pt] (char) {#1};}}

\begin{document}

\title{ZT-SDN: An ML-powered Zero-Trust Architecture for Software-Defined Networks}

\author{Charalampos Katsis}
\email{ckatsis@purdue.edu}
\author{Elisa Bertino}
\email{bertino@purdue.edu}
\affiliation{%
  \institution{Purdue University}
  \city{West Lafayette}
  \state{Indiana}
  \country{USA}
}

\renewcommand{\shortauthors}{Katsis et al.}

\begin{abstract}

Zero Trust (ZT) is a security paradigm aiming to curtail an attacker's lateral movements within a network by implementing least-privilege and per-request access control policies. However, its widespread adoption is hindered by the difficulty of generating proper rules due to the lack of detailed knowledge of communication requirements and the characteristic behaviors of communicating entities under benign conditions. Consequently, manual rule generation becomes cumbersome and error-prone. To address these problems, we propose \textit{ZT-SDN}, an automated framework for learning and enforcing network access control in Software-Defined Networks. ZT-SDN collects data from the underlying network and models the network ``transactions’’ performed by communicating entities as graphs. The nodes represent entities, while the directed edges represent transactions identified by different protocol stacks observed. It uses novel unsupervised learning approaches to extract transaction patterns directly from the network data, such as the allowed protocol stacks and port numbers and data transmission behavior. Finally, ZT-SDN uses an innovative approach to generate correct access control rules and infer strong associations between them, allowing proactive rule deployment in forwarding devices. We show the framework's efficacy in detecting abnormal network accesses and abuses of permitted flows in changing network conditions with real network datasets. Additionally, we showcase ZT-SDN's scalability and the network's performance when applied in an SDN environment.

\end{abstract}




\maketitle

\input{introduction}
\input{background}
\input{related_work}
\input{design}
\input{host_module}
\input{csm_module}
\input{arl_module}
\input{rtfsl_module}
\input{rgam_module}

\input{zt-gym}

\input{evaluation}

\input{conclusion}

\begin{acks}
The work reported in this paper has been supported by NSF under grant 2112471. The authors would like to thank Sumedh Nitin Joshi for his help in implementing the ZT-Gym.
\end{acks}

\bibliographystyle{ACM-Reference-Format}
\bibliography{sample-base}

\end{document}

%% file: introduction.tex
\section{Introduction}

Zero Trust (ZT)~\cite{stafford2020zero} is a security paradigm that can significantly enhance network security. With ZT, the traditional perimeter-based security model is replaced by a rigorous approach that focuses on authenticating and authorizing every access attempt, regardless of whether it originates inside or outside the network. By strictly controlling access and adhering to a \textit{``need-only''} basis, ZT ensures that only authorized parties can access critical resources and services in the least privileged manner. Such fine-grained control mitigates the risks of unauthorized access and curtails lateral movement by attackers. It reduces the impact of compromised entities, as attackers would have limited access to sensitive resources even if the network perimeter is breached.

\noindent{\bf Problem and Scope.} Enforcing strict network access control poses a significant problem for several reasons. First, achieving least privileged network access requires a comprehensive understanding of the communication requirements of every network component, including applications, IoT devices, and services~\cite{katsis2022neutron, nelson2010margrave, casado2007ethane, tongaonkar2007inferring, katsis2021can}. Second, once these requirements are identified, network administrators are tasked with manually generating appropriate flow rules to permit authorized traffic within the network. This process is highly susceptible to errors, and even the smallest oversight could lead to a chain of failures~\cite{katsis2022neutron}.  Last, often, there is no understanding of the benign behavior of applications utilizing those permissions during data transmission, such as the typical structure of exchanged packets and data exchange patterns. \textit{The goal of this work is thus to design an end-to-end automated framework for the learning, generation, deployment, and monitoring of ZT policies in network systems.}

We cast our design in the context of Software-Defined Networks (SDN)~\cite{kreutz2014software} as it is an effective framework for facilitating efficient solutions based on ZT principles. The reason is that SDN employs a centralized controller in the control plane to manage communication requests from data plane hosts and services. This centralized controller has the authority to make decisions for handling these communications, such as granting or denying access based on the established ZT policies or how the data should be routed in the data plane. By leveraging this centralized control, SDN enables a granular and adaptive approach to implement ZT, allowing for enhanced security and fine-grained control over network communications.

\noindent{\bf Challenges.} The design of our framework requires addressing the following challenges: \textbf{(C1)} The communication requirements of the network components are often not provided. Relevant information includes network-related intricacies, such as the allowed communication protocol stacks (e.g., \verb|ETHERNET_IP_TCP|). 
\textbf{(C2)} The operation of network components during benign conditions is not known in advance, rendering it impractical to effectively identify anomalous behaviors abusing permitted flows. That is, \textbf{C2(a)} the typical format of the messages and \textbf{C2(b)} the typical behavior of a particular component during data transmission are not often known.
\textbf{(C3)} The correct access control rules allowing the network components to complete their missions successfully are often unknown (e.g., allow bidirectional TCP and UDP communications at port 5555). 

\noindent{\bf Our Approach.} To address those challenges, we propose \textit{ZT-SDN}, a ZT framework that learns the benign communication patterns and behaviors of communicating entities (i.e., users, applications, services) from the underlying network and automatically generates fine-grained access control rules using unsupervised machine learning (ML) techniques while also understanding how those rules should be used in benign conditions.

ZT-SDN collects and analyzes data about the communications and patterns of every communicating entity. The analysis is carried out by identifying which entities use the network and how. This is achieved by collecting the network traffic and the network-related actions performed by applications on the hosts. 
ZT-SDN then constructs a \textit{communication requirements graph} where the nodes represent the communicating entities, and the edges represent the different network ``transactions'' performed. The network transactions are represented by the protocol stack of the exchanged network packets. Each transaction direction can be thought of as network access (addressing \textbf{C1}). Therefore, the key idea is to extract the patterns from those accesses. We extract the patterns of the network transactions in two forms: (1) the structure of packets involved in a transaction (e.g., protocol stacks used, port numbers) and (2) the characteristic behavior of data transmission. Those patterns are obtained using ML models and trained using datasets specific to each entity performing transactions. For extracting the packet structure, we use unsupervised artificial neural networks (ANN)~\cite{tschannen2018recent}. This allows ZT-SDN to determine the access patterns of benign data originating from a particular source in the graph, thereby enabling the identification of anomalous accesses that deviate from the learned distribution (addressing \textbf{C2(a)}). To extract the behavior of data transmission, ZT-SDN observes the pattern of data transmissions on ongoing connections using a novel time-series approach. The idea is to measure how far (or close) the observed patterns are to the learned patterns (addressing \textbf{C2(b)}). Those ML approaches allow for identifying deviant behavior and pinpointing the specific entity causing the issue.  
Deviant behavior may either occur due to attacks or abnormal yet benign behavior. A typical example of the latter case is two applications with different transmission patterns, yet both of them are benign applications. We demonstrate that our approach can effectively identify the abnormality in both attack and abnormal benign patterns while exhibiting a reasonable tolerance against changing network conditions, including reduced bandwidth, link delays, and jitter.

The quality of the extracted patterns is intricately linked to the duration of the training process. However, given the diverse complexities of different applications, determining the optimal amount of data required for each can be challenging. To address this issue, ZT-SDN incorporates user-provided training heuristics as part of the solution. Such heuristics include the minimum training time and the desired performance on unseen benign training data. Then, ZT-SDN employs an iterative learning approach, continuously updating the models until the performance requirements are achieved. 
The training is then terminated or resumed until those requirements are met.

Finally, ZT-SDN analyzes the network patterns and generates \textit{correct} flow rules, which are installed in the data plane devices (i.e., network switches), allowing the permitted traffic to pass while restricting transactions not encountered during training. We introduce an innovative approach wherein ZT-SDN  autonomously infers the appropriate protocol stack features (i.e., predicates) for rule construction and their respective values. This process involves measuring the correlations between protocol header-value pairs through association rule mining techniques. Subsequently, ZT-SDN identifies the most extensive set of feature-value pairs that can collectively constitute the rule with the highest association measure (e.g., 100\%). As a result, the constructed rules are correct, as each header-value pair strongly associates with the rest of the pairs in the dataset (addressing \textbf{C3}).
Our framework not only generates the rules but also finds strong associations between them. The core concept revolves around identifying which of the generated rules apply to the traffic transmitted within a specific time window and how often. 
For example, as TCP is a bidirectional protocol, the forward and backward direction rules will be automatically classified as strongly associated, thus capturing the intricacies of network protocols. ZT-SDN leverages this knowledge by deploying strongly associated rules together in the data plane. Our results show that such an approach reduces the number of interaction roundtrips between the control and data planes, thus improving efficiency.

ZT-SDN's training process is designed to be performed directly from the underlying network, assuming all collected data are benign. However, this assumption may be impractical in many network scenarios where adversarial absence cannot be guaranteed. To address this concern, we have developed \textit{ZT-Gym}, an alternative approach for facilitating data generation and model training offline before integration into the SDN infrastructure. ZT-Gym operates as a Linux-based controlled environment where relevant applications or services are deployed and executed, generating datasets for offline model training.

\noindent{\bf Novelty.} To the best of our knowledge, ZT-SDN is the first end-to-end ZT pipeline for automated learning, enforcement, and monitoring of access control policies in the context of SDN.

\noindent{\bf Contributions.} The paper makes the following contributions:
\begin{itemize}
    \item ZT-SDN, a novel end-to-end ZT architecture for SDN networks.
    
    \item Automated learning processes supported by datasets collected from the underlying network infrastructure without requiring prior knowledge about the communication requirements of applications or services.
    
    \item ZT-Gym, a virtualized environment for deploying software applications and offline dataset generation and model training.
    
    \item Techniques for profiling different aspects of communications executed by different applications using unsupervised ML models and making access control decisions.
    
    \item A novel technique for automatic flow rule generation and identification of strong rule associations.
\end{itemize}

%% file: background.tex
\section{Background}

% Flow tables
In SDN, the network policies are installed and enforced in the network in the form of \textit{flow rules} where each rule consists of a set of matching predicates (e.g., source/destination addresses, protocol, port numbers) and corresponding actions (e.g., forward, drop, modify) to be performed on matching packets. The flow rules are communicated from the network controller to the forwarding devices (i.e., switches) via the \textit{southbound channel}, which uses a communication protocol that those devices support (e.g., OpenFlow~\cite{mckeown2008openflow} and ForCES~\cite{yang2004forwarding}). Once a switch receives the rules, it installs them in its \textit{flow table}. Each rule within a switch is assigned a unique identifier set by the network controller, known as a \textit{cookie} value according to the OpenFlow specification. This unique identifier, set by the network controller, facilitates actions ordered by the controller, such as rule updates or revocations.

% Reactive and proactive forwarding
Typically, the forwarding rules are installed according to a \textit{reactive forwarding} approach. In this approach, the controller installs a rule on a switch as a result of a packet receipt at an input port of the switch. The switch first checks its flow table to see if there is any matching policy for handling the packet. The rules in the switch are matched in descending order of \textit{priority}. The rule with the lowest priority is typically installed by the network controller (ONOS) and instructs the switch to forward the packet to the controller encapsulated in a \verb|OFPT_PACKET_IN| message~\cite{openflowSwitch}. We define this as the \textit{default rule}. This message includes the default rule's cookie (responsible for triggering the \verb|PACKET_IN|) and the switch's identifier, providing essential information to the controller. 
The controller checks the request and makes a decision. If the packet should be forwarded to its destination, the controller replies with a \verb|OFPT_PACKET_OUT| message~\cite{openflowSwitch} to the switch indicating how the packet should be dispatched (e.g., forward packet to output port 2). Typically, after the latter message, the controller sends to the switch a \verb|OFPT_FLOW_MOD| message~\cite{openflowSwitch}, which includes a rule that the switch needs to install in its forwarding table.
The controller may set a rule timer where the switch revokes the rule automatically after a period of inactivity.
The controller may also set the \textit{hard} and \textit{idle expiration timers}, which instruct the switch to revoke the rule after the specified time. The hard timer revokes the rule after a specified number of seconds, whereas the idle timer revokes the rule only if there is no packet matching the rule for the specified number of seconds. As long as a rule has not expired, future packets matching the rule are directly handled as per the controller-specified treatment. Thus, the switch does not need to consolidate with the controller again, which saves a significant amount of time. 
The controller can also install forwarding rules \textit{proactively}. Such an approach reduces the performance overhead imposed when the switch has to hold the packet transmission until a decision is received from the control plane~\cite{isyaku2020software}. However, one must know the policies that must be installed in advance, which may not be the case.

The forwarding decisions in the control plane are handled by \textit{network applications}. The controller exposes control plane functionality to network applications via its \textit{northbound channel}. This functionality is typically exposed as Application Programming Interfaces (APIs) or REST APIs. An application registers to listen to specific events (e.g., \verb|PACKET_IN|) and may provide the logic that handles those events. 

%% file: related_work.tex
\section{Related Work}

\noindent\textbf{Network access control.} Several approaches have been proposed for enforcing access control in organizational networks~\cite{nayak2009resonance, katsis2022neutron, casado2007ethane, anjum2022removing, anjum2023msnetviews, csikor2022zerodns, vanickis2018access}. Nayak et al.~\cite{nayak2009resonance} proposed Resonance, a framework for the specification and enforcement of access control policies. 
Katsis et al.~\cite{katsis2022neutron} proposed NEUTRON, a graph-based framework for defining and enforcing least privilege access control policies. 
Csikor et al.~\cite{csikor2022zerodns} proposed an approach to implement ZT on DNS infrastructure, thereby controlling client domain name resolutions based on their identity and privileges. Anjum et al.~\cite{anjum2022removing} proposed a framework supporting the definition of least privilege access control policies for SDN. 
Vanickis et al.~\cite{vanickis2018access} proposed high-level framework, FURZE, for ZT networking. The framework is composed of multiple components involving policy authoring using FURZE language for generic AC policies and firewall rules. 
All those approaches have many limitations: (1) they assume that the network-wide security policies are provided by some administrator; (2) they assume policies are correct and detailed, so the generated rules allow ``need only''-based communication; (3) they do not capture behavioral aspects of the communications such as the message or data transmission patterns which are essential to understand how the permissions are used in benign conditions; (4) the predicates used in the generated rules are limited to administrator-specified fields (e.g., source and destination addresses and ports). All these limitations are addressed by ZT-SDN in an automatic manner. 

\noindent\textbf{Rule mining.} Golnabi et al.~\cite{golnabi2006analysis} proposed an approach to mine generalized firewall policy rules from network traffic logs based on rule frequency analysis. 
Apiletti et al.~\cite{apiletti2009characterizing} proposed the NetMine framework to extract generalized rules from network traffic data using association rule mining. 
Those approaches have the following limitations: (1) The rule extraction is limited to user-specified predicates or user input; (2) the generalization process is not designed with least-privilege network access in mind, which results in rules allowing a potentially wide range of hosts to access network resources. ZT-SDN addresses those limitations. It analyzes the packets transmitted between different hosts separately. Then, it uses a novel rule generation approach that automatically extracts the predicates and uses them for constructing rules with high confidence and support. Finally, ZT-SDN generates rules specific to the communicating entities, restricting access to resources not specified in the CR graph.

\noindent\textbf{Anomaly detection in SDN flows.} Sayed et al.~\cite{el2022flow} proposed a supervised learning method based on Long Short Term Memory and AE to detect distributed denial of service (DDoS) attacks. Peng et al.~\cite{peng2018detection} proposed another supervised approach based on the KNN algorithm for detecting DDoS attacks. 
Da Silva et al.~\cite{da2016atlantic} developed the ATLANTIC framework, which identifies flow anomalies based on entropy analysis and a supervised algorithm based on support vector machine (SVM). 
Garg et al.~\cite{garg2019hybrid} proposed a supervised deep learning-based anomaly detection scheme for suspicious flow detection in the context of social multimedia. 
Additionally, Nanda et al.~\cite{nanda2016predicting} evaluated various supervised ML algorithms
trained on historical network attack data to predict potential malicious flows. Unsupervised approaches have also been proposed to detect anomalies based on techniques such as entropy analysis, clustering and isolation forests~\cite{abou2021novel, carvalho2018ecosystem, scaranti2022unsupervised}.

In contrast, our ZT-SDN framework offers a novel approach, fundamentally different from prior works. Rather than focusing solely on detecting ongoing attacks, we aim to prevent abuse of network permissions by entities by understanding how those permissions are used in benign conditions. Our approach, specifically the RTFSL module, distinguishes itself in three key aspects: (1) it operates entirely in an unsupervised manner; (2) it tailors the training to the specific entities using the network permissions, thereby understanding how the network permissions are typically used by the communicating entities; and (3) it detects anomalies in granted flow permissions using time-series modeling to assess pattern similarities between observed and learned patterns (e.g., sending too fast/slow or too much/few data).

\noindent\textbf{Relation to NIDS Systems.} While ZT-SDN and Network Intrusion Detection Systems (NIDS) share complementary goals in network security, critical distinctions set them apart. NIDS technologies function at the data plane, continuously monitoring traffic streams for anomalies based on diverse indicators, such as connection resets, SYN attempts, and traffic patterns~\cite{lunardi2022arcade, mirsky2018kitsune, holland2021new}. This monitoring allows NIDS to classify traffic and detect attack signatures accurately. However, NIDS does not typically perform access control enforcement; it is typically deployed at specific network ingress or egress points and thus cannot comprehensively mediate all traffic between every network entity. Deploying NIDS systems to every access point across all subnets would lead to prohibitive costs and complex management challenges. Additionally, relocating NIDS functionality to the control plane is impractical due to the overwhelming volume of data plane traffic, which would overwhelm the control plane resources.

NIDS provides alerts on detected anomalies but does not handle access control decisions, meaning it cannot assess whether a communication should be permitted or denied. Furthermore, the core objective of NIDS is to detect attacks rather than manage deviations in network permissions assigned to specific components, making it unable to distinguish subtle variations between legitimate traffic flows. In contrast, ZT-SDN operates within the control plane, enabling it to dynamically enforce network-wide access control policies. ZT-SDN actively decides whether packets should be allowed or blocked, and it can analyze flow statistics in real time, aligning with OpenFlow specifications, to detect deviations in permitted network behavior. 
Thus, while ZT-SDN and NIDS both contribute to network security, they cannot be directly compared or replace one another.

%% file: design.tex
\section{Design of ZT-SDN}

In this section, we first present an overview of the ZT-SDN architecture, followed by an outline of the threat model and assumptions. 

\begin{figure}[htbp]
  \centering
  \includegraphics[width=.6\linewidth]{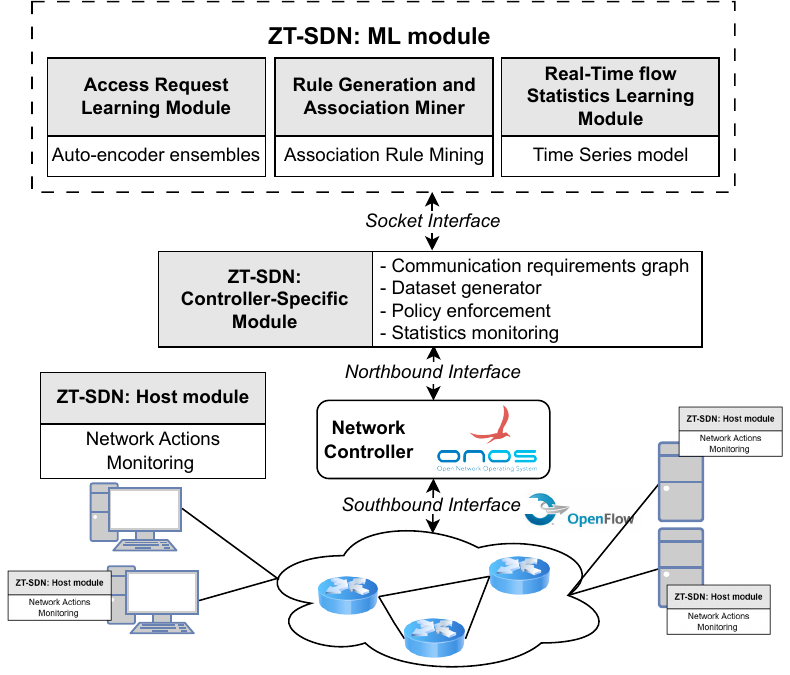}
  \caption{The ZT-SDN architecture.}
    \label{fig:architecture}
\end{figure}

\subsection{Architecture Overview}
\label{sec:arch-overview}

ZT-SDN has three modules (see Figure~\ref{fig:architecture}): the host module (HM), the controller-specific module (CSM), and the machine learning (ML) module.
The HM provides information about which applications on the hosts require network access (addressing \textbf{C1}). The CSM is a network application that consumes the northbound interface exposed by the network controller (i.e., ONOS). Thus, the execution of the network application has to be adjusted to the underlying controller. This module performs a different set of operations depending on its mode:

\noindent\textbf{Training mode:} It identifies the communicating entities by analyzing network traffic from the data plane. It also leverages information from the HM instances to map specific traffic streams in the data plane to the corresponding applications running on the hosts. This mapping facilitates the extraction of a \textit{communication requirements (CR)} graph, representing different sets of communications (edges) between the identified entities (nodes). For each edge in the graph, CSM generates datasets used by the ML module for training the models and extracting the communication patterns (addressing \textbf{C1, C2}). Once the models have been trained, the module transitions into enforcement mode.

\noindent\textbf{Enforcement mode:} In this mode, the CSM uses the ML module to make network access control decisions. It also constructs the rules generated by the ML module and installs them at the network's appropriate switches (i.e., policy enforcement points -- PEPs). It retrieves flow statistics associated with the installed flow rules in real time and provides them to the ML module.

The ML module is a controller-independent module, thus allowing interoperability across different SDN controllers. Like the CSM module, it operates in two modes:
    
\noindent\textbf{Training mode:} During this mode, the ML module processes the data plane traffic datasets. 
% provided by the CSM. 
Utilizing unsupervised learning techniques, it extracts the benign communication patterns modeled by the CR graph and derives corresponding flow rules (addressing \textbf{C1, C2, C3}). 

\noindent\textbf{Enforcement mode:} In this mode, the trained models authorize network access requests from hosts in the data plane based on temporal factors and packet headers. Furthermore, it continuously analyzes flow statistics of ongoing communications to identify potential deviant behaviors.
% \end{itemize}

\subsection{Threat Model and Assumptions}

ZT-SDN operates under the assumption that no prior knowledge is available regarding the applications running on the hosts in a network. Additionally, we assume that there is an authentication process for the host machines before the training process commences. This is essential, especially in cases where the entity network identifiers (e.g., IP or MAC) change over time. Therefore, endpoint identity is used for training and access request authorization.

Throughout the ML training phase, it is crucial that the host applications running are not compromised. However, once the model is trained and the ML and CSM modules switch to the enforcement mode, an adversary may compromise host applications or services running inside or outside the network. Finally, we assume that the control plane operations (i.e., network controller) and the data plane forwarding devices (i.e., the PEPs) are never compromised at any point in time. 

Nevertheless, the assumption that the host applications are not compromised during the training phase may be overly stringent for practical adoption in some organizations, particularly in scenarios where network administrators cannot guarantee the absence of abnormal network activity, even during the training phase. To address this concern, we develop an \textit{offline approach} to dataset generation and model training. In this approach, applications are deployed and executed within a controlled ``gym environment,'' referred to as \textit{ZT-Gym} –- a virtualized setting where the application(s) under scrutiny run. 
Network traces collected from \textit{ZT-Gym} are then used for training the models for deployment in the network.
We provide technical details of this alternative approach in Section~\ref{sec:zt-gym}. The approach discussed in the rest of the section performs the training phase directly from the network.

%% file: host_module.tex
\section{Host Module (HM)}
\label{sec:hm}

The HM is installed on the host systems (i.e., endpoints). It is a lightweight Python program that continuously interacts with the underlying OS to detect network-related events by system processes. Such events include port bindings and data transmissions. It collects and aggregates essential information associated with these events by interacting with the OS. This includes the timestamp of event detection, applications binding OS ports, identification of applications transmitting data, the transport-layer protocol utilized by the applications, and destination information such as IP addresses and ports.  
It then transmits this information to the CSM, allowing for the association of the network flows with the host applications.

\noindent\textbf{Support for Embedded Systems.} The HM’s installation is not required on embedded systems, including the Internet of Things (IoT) and operational technology control devices. Given that these devices execute highly specific tasks, they can be treated as standalone applications.

%% file: csm_module.tex
\section{The Controller-Specific Module (CSM)}
\label{sec:csm}

The CSM builds upon ONOS's reactive forwarding application, known as Fwd, to manage the \verb|PACKET_IN|, \verb|PACKET_OUT|, and \verb|FLOW_MOD| operations, which facilitate effective communication between entities in the data plane. The initial step involves the CSM registering with the network controller via the northbound interface and requesting to handle any \verb|PACKET_IN| messages from the data plane.

\noindent\textbf{Training mode.} When a packet is sent from a host in the data plane, the switch sends it to the CSM encapsulated in \verb|PACKET_IN| message. This happens when there is no existing flow rule permitting communication, essentially indicating a network access or communication request from a source to a destination. Subsequently, the CSM extracts the source and destination addresses from the packet, along with the packet's protocol stack (e.g., \verb|ETHERNET_ARP| or \verb|802.11_IP_TCP|). 
 
The CSM identifies the communicating hosts using the extracted information and determines the applications involved through HM updates. It then utilizes this knowledge to create the CR graph. In such a graph, each node represents an application running on a host, encompassing both client-side applications and network services. The directed edges between the nodes represent the identified protocol stack used for communication. Essentially, each edge in the graph represents a necessary network access, and it is defined by the source and sink nodes and the protocol stack of the transmitted packets over that edge. The CR graph represents the essential flows required for applications to perform their intended functions successfully. Additionally, it provides insights into the endpoints, such as IP addresses and domain names, with which applications typically communicate.

\begin{figure}[h]
  \centering
  \includegraphics[width=.3\linewidth]{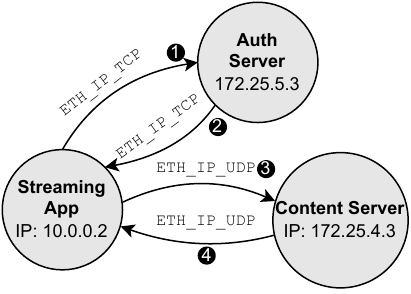}
  \caption{An example of a simple communication requirements graph.}
    \label{fig:com-req}
\end{figure}

Figure~\ref{fig:com-req} shows a simple example of a communication requirements graph.
The example illustrates a streaming application that uses two protocol stacks: \verb|ETHERNET| \verb|_IP_TCP| (flows \circled{1}--\circled{2}) for the authentication procedure, and  \verb|ETHERNET| \verb|_IP_UDP| (flows \circled{3}--\circled{4}) for content consumption.

For every discovered edge in the CR graph, the CSM generates two types of datasets: (1) packet datasets and (2) flow statistics datasets. The packet datasets capture all transmitted packets associated with a specific edge in the graph. For instance, if an application communicates with TCP and UDP protocols, the CSM will generate two individual datasets for the application: one for the TCP packets and one for the UDP packets. The CSM currently supports various protocols, such as ARP, IP, IGMP, ICMP, TCP, and UDP. Each entry in the dataset contains temporal information (i.e., day, time) and the parsed packet header, which includes protocol-level details such as addresses, VLAN ID, frame types, ports, and flags.

\begin{figure}[h]
  \centering
  \includegraphics[width=.5\linewidth]{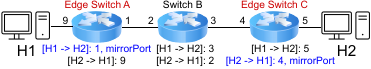}
  \caption{
    Rule deployment for traffic collection and forwarding. The notation is [matching packet headers]:[List of forwarding interfaces].
    \label{fig:data_collection}
    }
\end{figure}

The CSM then sends a \verb|PACKET_OUT| command to forward the packet to its intended destination. The CSM installs a custom rule using the source and destination addresses, ports, and the transport layer protocol. This rule ensures that subsequent packets related to the same flow can be efficiently forwarded to their destinations. The action set of the rule is configured to forward the packet through a designated egress port determined by ONOS' topology service~\cite{onosTopoService} at the time of rule installation. For the \textit{edge} switch\footnote{Edge switch is the switch where a host attaches to the network.}, an additional action is configured in the flow rule. This additional action instructs the edge switch to send a copy of matching packets to a \textit{mirroring} port\footnote{Port mirroring is typically supported by hardware for efficiency~\cite{ciscoPortMirroring}.} (see Figure~\ref{fig:data_collection}).
The mirroring port of each edge switch is connected to an end system that passively collects data, such as the controller or organization-owned hosts\footnote{There are several efficient approaches for packet capturing at least 100 Gbps, such as \cite{emmerich2017flowscope}. This choice doesn't impact ZT-SDN.}. 
While the CSM can collect traffic data passively at the controller, this may consume significant bandwidth and processing power, particularly with numerous hosts connected to an edge switch. This can slow down control and data plane operations. Therefore, a more scalable alternative is for the CSM to obtain traffic data separately from the mirroring systems.

Upon installing the corresponding rule in a switch’s flow table, the CSM utilizes the ONOS flow statistics service~\cite{kreutz2014software} to generate datasets for different CR graph edges. For each flow direction, the CSM obtains flow statistics for the edge switch connected to the source host. For instance, in Figure~\ref{fig:data_collection}, the flow H1$\rightarrow$H2 is queried at switch A, while the flow H2$\rightarrow$H1 at switch C. 

The flow statistics that can be collected by OpenFlow switches are predefined by the OpenFlow specification~\cite{openflowSwitch}. The available statistics for each flow include the total number of packets and bytes transmitted since the installation of the flow rule until the time of the query. Thus ZT-SDN has access to those statistics only. The queries are performed by the controller at predetermined intervals of $f$ seconds. 
At the time of the query $t$, the number of transmitted packets and bytes are computed as follows: 
\begin{equation} \label{eq:packet_counting}
 \#packets_t = \sum_{rule\in edge.rules}{rule.packets_t}   
\end{equation}
and 
\begin{equation} \label{eq:byte_counting}
 \#bytes_t = \sum_{rule\in edge.rules}{rule.bytes_t}   
\end{equation}
In essence, the number of packets (or bytes) at $t$ is computed as the sum of packets (or bytes) for each installed flow rule associated with the specific edge in the CR graph. Each entry in the dataset comprises the query time ($t$) along with the total number of packets ($\#packets$) and bytes ($\#bytes$) reported during that query.

ZT-SDN performs data collection operations at edge switches instead of core switches because (1) edge switches mediate the communication even in multi-path packet propagation, and (2) edge switches typically handle less traffic load than core switches, allowing for more scalable data collection.

\noindent\textbf{Enforcement mode.} 
The CSM parses the \verb|PACKET_IN| message (similar to the training mode) and performs a query on the CR graph to determine if the specific communication has been previously modeled during the training phase. If it finds a match in the graph, the packet is then forwarded to the access request learning (ARL) module for verification against the learned distribution of packet headers. Upon successful verification, the CSM receives the flow rules generated by the rule generator and association miner (RGAM) module and proactively deploys them across all switches between the source and the destination. The path for rule deployment is determined by ONOS at the time of rule deployment. Similar to the training mode, the action set for the rules is set to ``forward'', and the egress ports are specified based on the path determined by ONOS. However, if the communication has not been modeled during training in the CR graph or if the packet does not belong to the learned distribution (as determined by ARL), the communication request is denied. ZT-SDN allows network administrators to set hard and idle flow timers for the deployed rules. By default, it sets an idle timer at 10 seconds, so inactive rules are retracted from the data plane.

The ZT-SDN system allows network administrators to dictate if copies of subsequent packets, for either all or specific edges in the CR graph, should be forwarded to the controller during enforcement mode, and this can even be set for a specific time duration. This decision, however, involves balancing security and network performance. Forwarding packet copies to the controller enables the ARL module to conduct more thorough security checks, thereby enhancing network security. Conversely, this method increases bandwidth usage, negatively impacting network performance. Therefore, administrators need to consider these aspects and decide what best aligns with their specific security and performance needs.

Following the deployment of the rules, the CSM continues to query the statistics of the installed flow rules in the edge switches (similar to the training period), allowing for real-time flow monitoring. The collected statistical information is passed to the real-time flow statistics learning (RTFSL) module, which actively checks for any deviations in the usage of allowed network flows. If any deviation is detected, the CSM ends the communication between the endpoints by removing the corresponding flow rules from the data plane, effectively cutting off the communication. 

\noindent\textbf{Network Overheads.} The processing of \verb|PACKET_IN| requests in ZT-SDN increases due to the CSM processing and the access control decision made by the ARL module. However, as demonstrated in Section~\ref{sec:scalability}, this increase in processing is offset by the proactive rule deployment approach. The RTFSL module does not cause network overhead as the ONOS controller obtains flow statistics from all network switches at a predetermined frequency. Thus, it utilizes the statistics that have already been collected. Lastly, rule generation is done once during training, so it does not add overhead during enforcement.

%% file: arl_module.tex
\section{The Access Request Learning module (ARL)}
\label{sec:arl}

The module's objective is to determine whether to grant or deny a communication request originating from an entity in the data plane by evaluating the packet contained in a \verb|PACKET_IN| request. To achieve this, the module must learn the allowed values for the different protocol stack headers typically used for the specific edge in the CR graph, such as the permissible port numbers, time-to-live values, typical header and packet lengths.

\begin{figure}[h]
  \centering
  \includegraphics[width=.9\linewidth]{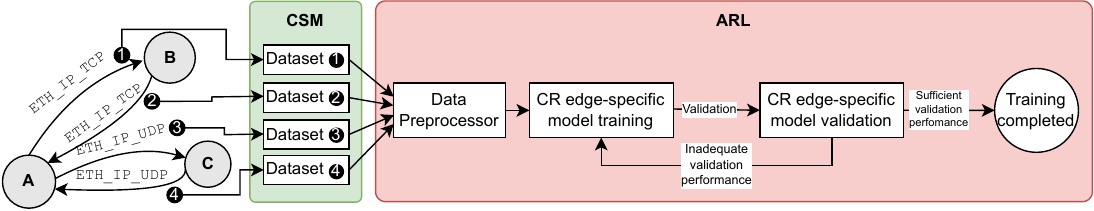}
  \caption{
    The model training procedure for each CR graph edge.
    \label{fig:arl-training}
    }
\end{figure}

\noindent \textbf{Training.} \verb|PACKET_IN| requests may contain a packet that can be exchanged at any point in time between entities, as connections may be interrupted and later resumed from where they left off. Therefore, training solely on the initial packets of an edge is insufficient; the module must be able to recognize any benign exchanged packet. The ARL receives traffic datasets generated for each application from the CSM, with each dataset corresponding to a different protocol stack. Figure~\ref{fig:arl-training} shows the ARL training process. Each sample in these datasets contains the day and time of packet receipt along with the protocol header values. 

To prepare the data, we first apply protocol-specific processing, which involves feature transformations and deletions. For instance, to preserve the semantic meaning of the TCP protocol flags, we one-hot encode the TCP flags to differentiate the set flags for each packet. Protocol features with no meaningful values for the access control learning process are removed, such as acknowledgment values and checksums. Then, standard scaling is applied, which removes the mean and scales the data to unit variance. We then a model comprised of autoencoders (AEs) ensembles for each CR edge to capture the distribution of packet headers in that edge. An AE is a neural network architecture used for unsupervised learning, primarily designed to learn efficient representations of input data by reconstructing the original input from a compressed intermediate representation. The goal of an AE is to minimize the reconstruction error—the difference between the original input and the decoder output—thus learning essential features and patterns of the input data in the latent space.

We specifically use the KitNet model~\cite{mirsky2018kitsune}, which is a lightweight, highly efficient CPU-based model built on AE ensembles. KitNet uses a feature mapper to create smaller clusters of features, with each cluster fed to a separate AE in the ensemble. This allows the AE ensemble to capture and learn the distinguishing aspects of normal network behavior, enabling the model to effectively identify deviations and anomalies.

KitNet operates in two modes: training and execution. Initially, it requires the specification of the number of training examples. Once this number of samples has been observed, KitNet transitions to the execution mode. In this mode, the model calculates the reconstruction error for all subsequent samples. However, this approach poses a significant challenge, as determining the precise number of training samples needed beforehand can be difficult. First, different applications vary in complexity, necessitating a unique number of training instances for each application-specific model. Second, the model must capture the temporal factors associated with the access request, such as the time of the request. To address these issues, we have modified the algorithm's design by allowing network administrators to provide heuristics on the desirable model performance on both seen and unseen training data. 
We perform the following modifications:
\begin{enumerate}
    \item Introduction of the minimum training time. This refers to the minimum duration during which an application is expected to demonstrate its typical behavior associated with temporal factors. 

    \item Introduction of the minimum number of training samples. We establish a minimum threshold for the number of training samples required to adequately train the model (e.g., 20K samples). 

    \item Incorporation of a validation period. A designated time span is introduced to assess the model's performance on unseen benign data. 
    
    \item Introduction of a maximum reconstruction error threshold. During the validation period, we define a maximum reconstruction error threshold. If the average reconstruction error during the validation period falls below this threshold, indicating satisfactory performance, the model proceeds to the execution mode. Otherwise, it resumes training to improve its accuracy.
\end{enumerate}
Therefore, the model training becomes an iterative process where the model is updated until the user-provided performance requirements are met. The model observes data for a minimum duration and a minimum number of training samples, and then its performance during the validation period determines the transition to the execution mode.

\noindent\textbf{Enforcement:} The packet is extracted from the \verb|PACKET_IN| message, and the endpoint addresses are used to identify the corresponding edge in the CR graph. The packet is then preprocessed and passed through the scaler, followed by the inference of the KitNet model trained on that edge, which either allows or denies the communication.

%% file: rtfsl_module.tex
\section{The Real Time Flow Statistics Learning Module (RTFSL)}
\label{sec:rtfsl}

When an access request is granted based on the CR graph and the ARL module, a flow rule or policy is installed, allowing the traffic to proceed to its intended destination. However, from a ZT perspective, we want to ensure that the granted network permissions are used as per the learned transmission patterns. For instance, a resource that typically transmits very little data in an active flow starts transmitting rapidly or at the same rate with very different payload sizes. 
Therefore, the RTFSL module addresses two critical requirements: 
(1) learning the normal flow usage permitted by the access rules for the modeled entities in the CR graph and (2) detecting deviations from the learned communication patterns and terminating the flows.

\noindent \textbf{Training:}
The RTFSL module uses the flow statistics datasets generated by the CSM for training. The collected statistics are limited to the ones supported by the Openflow specification~\cite{openflowSwitch}, ensuring the compatibility of our approach with the specification and, by extension, the vendor implementations. The training examples are essentially measurements of the transmitted packets and bytes transmitted over time. The data are sampled from a network switch at a predefined frequency, $f$ the total number of packets and bytes are computed as per equations \ref{eq:packet_counting} and \ref{eq:byte_counting}.
Thus, due to the data's semantic meaning and structure, we model it as a time series to capture the inherent dependencies and patterns present in time-dependent data. 

\begin{figure}[htp]
    \centering
    \begin{subfigure}[b]{0.45\textwidth}
        \centering
        \includegraphics[width=1.1\textwidth]{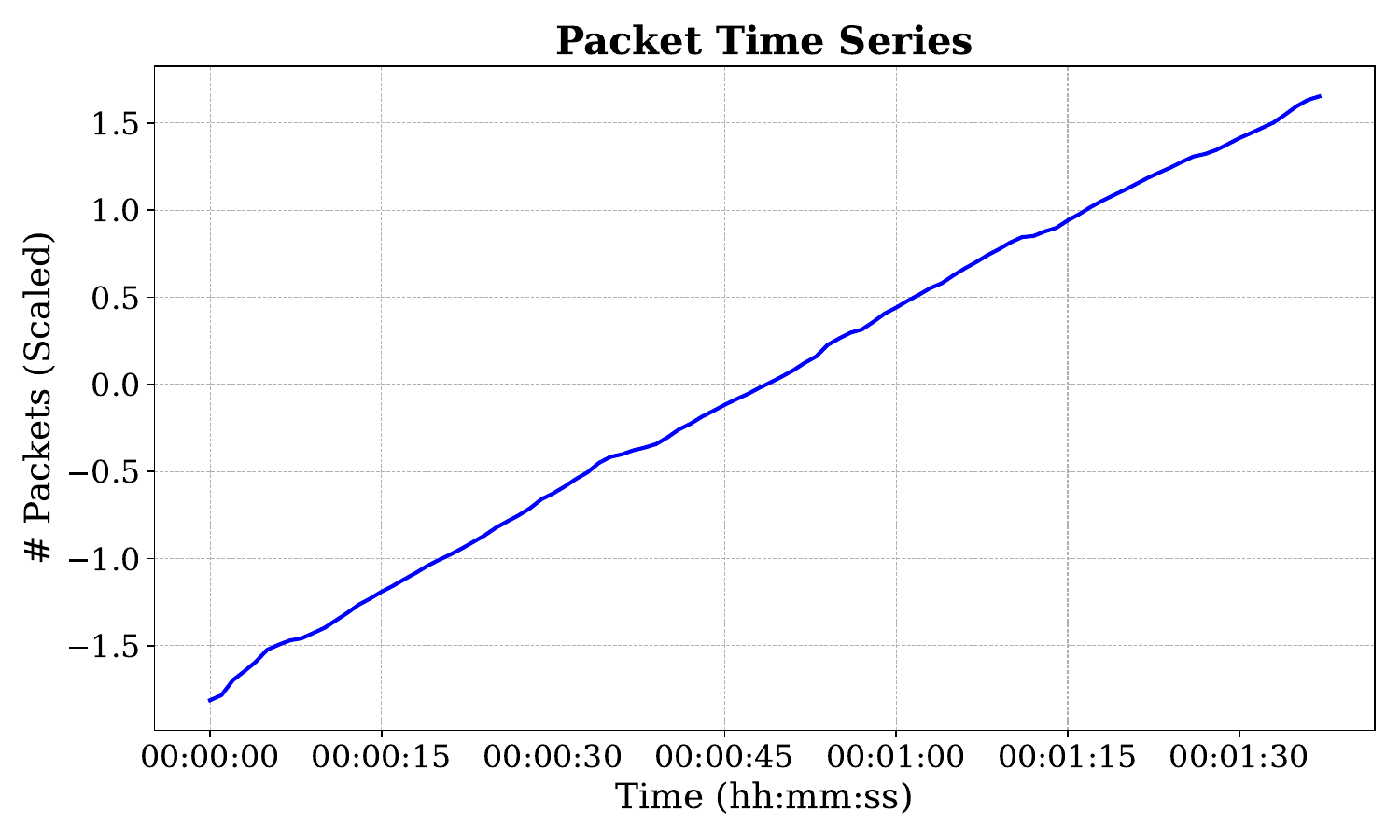}
        \caption{Packet Time Series (\( \{x_t\}_{t=1}^n \))}
        \label{fig:original_timeseries}
    \end{subfigure}
    \hfill
    \begin{subfigure}[b]{0.45\textwidth}
        \centering
        \includegraphics[width=1.1\textwidth]{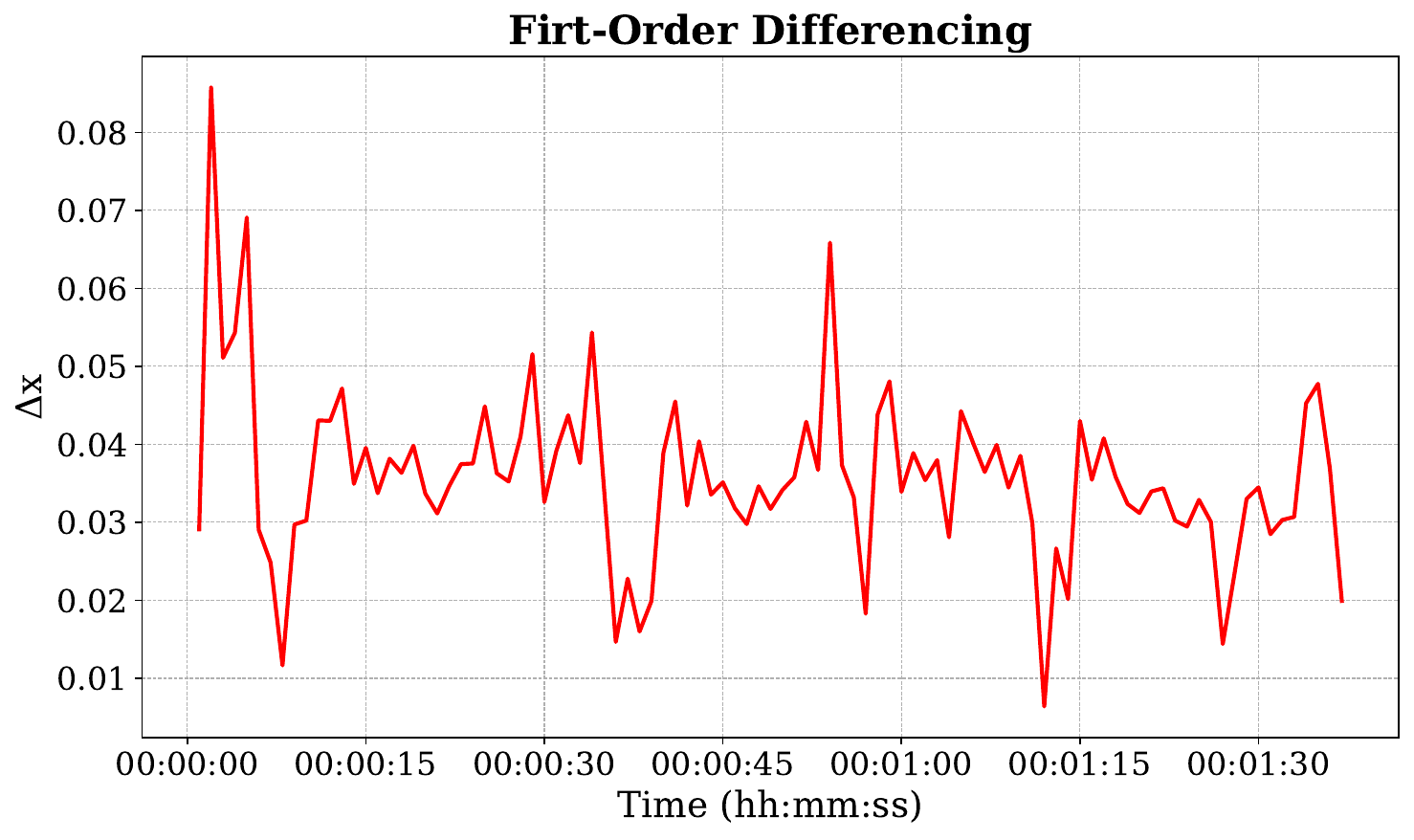}
        \caption{First-Order Differencing Time Series (\( \Delta x_t \))}
        \label{fig:first_order_difference}
    \end{subfigure}
    \caption{Transformation of the packet time series, (a), to the first order difference representation, (b).}
    \label{fig:timeseries_comparison}
\end{figure}

We first scale the training data of the CR edge using standard scaling, which removes the mean and scales to unit variance. We then remove the trend to reveal the data transmission patterns over the sampling time, as shown in Figure~\ref{fig:timeseries_comparison}. Figure~\ref{fig:original_timeseries} shows the cumulative packet transmission behavior of a real client machine transmitting to an HTTPS service from the MAWI 2024 traffic dataset~\cite{mawi_wide_dataset}.
To remove the trend, we apply first-order differencing to the data as follows: Given a time series \( \{x_t\}_{t=1}^n \), the first-order differencing is formally defined as: 
\begin{equation}
    \Delta x_t = x_t - x_{t-1}, \quad \text{for } t = 2, 3, \dots, n
\end{equation}
where \( \Delta x_t \) represents the first-order difference of the series at time \( t \). Figure~\ref{fig:first_order_difference} shows \( \Delta x_t \), which clearly shows the behavioral patterns between consecutive observation samples.

The idea is to collect long enough \( \Delta x_t \) time series from the CR edge such that future benign observations from the same edge can be matched to the \( \Delta x_t \) model with low error. We develop an approach to detect behavioral deviations in ongoing traffic flows (Algorithm~\ref{alg:traffic_pattern_match}) from the series \( \{ \Delta x_t\}_{t=2}^n \).  
We query the edge's traffic flows with the same sampling rate $f$ for some pre-determined time duration $d$. From the observation, we compute the first-order difference (line~\ref{alg:traffic_pattern_match:fod}) of the measurements. We perform dynamic time warping (DTW)~\cite{salvador2007toward}, which is a distance measure that quantifies the similarity between two-time series (line~\ref{alg:traffic_pattern_match:fdtw}). DTW considers the possible variations in the alignment and scaling of the time series, making it suitable for matching time series with different lengths and shapes. We use the Euclidean distance as the distance metric between the two time series. Our experiments with various time series shapes and sizes have shown that Euclidean distance is a good quantification metric for measuring similarity between them. Finally, we set a maximum distance threshold where patterns that exceed the threshold are added to the  \( \{ \Delta x_t\}_{t=2}^n \) series. We use the FastDTW implementation of DTW, which has a $O(n)$ time and space complexity, where $n$ is the length of the time series.

\begin{algorithm}
\caption{Traffic Pattern Similarity Checker}\label{alg:traffic_pattern_match}
\begin{algorithmic}[1]
\Require \( \{ \Delta x_t\}_{t=2}^n \), New Observations
\State $new\_obs \gets$ first\_order\_diff(New Observations)\label{alg:traffic_pattern_match:fod}
\State $min\_dist \gets \infty$
\For{$i$ in range(size(\( \{ \Delta x_t\}_{t=2}^n \)) - size($new\_obs) + 1$)}
\State $segment \gets  \{ \Delta x_t\}_{t=2}^n[i:i+size(new\_obs)]$
\State $dist \gets$ DTW($new\_obs, segment$)\label{alg:traffic_pattern_match:fdtw}
\If{$dist < min\_dist$}
\State $min\_dist \gets dist$
\State $min\_match\_segment \gets segment$
\EndIf
\EndFor
\end{algorithmic}
\end{algorithm}

We use four heuristics to stop the learning process:
\begin{enumerate}
    \item we use a predefined minimum data duration in which the model should be able to see all the potential traffic patterns.

    \item we use a minimum number of training samples in the \( \{ \Delta x_t\}_{t=2}^n \)

    \item we define a predefined validation period and a maximum Euclidean distance threshold to ensure the model works well on unseen benign data. 
\end{enumerate}

\noindent \textbf{Enforcement:} During model enforcement, the new observations are scaled and compared with the learned \( \{ \Delta x_t\}_{t=2}^n \) using Algorithm~\ref{alg:traffic_pattern_match}. We then define an anomaly threshold, which, if it exceeds the pattern, is recognized as anomalous.

\begin{figure}[h]
  \centering
  \includegraphics[width=.5\linewidth]{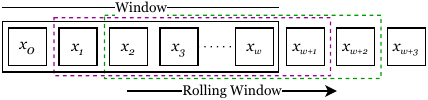}
  \caption{
    RTFSL rolling window.
    \label{fig:rtfsl-rolling-window}
    }
\end{figure}

We define a window size, $w$, which denotes the number of consecutive flow statistics samples, as shown in Figure~\ref{fig:rtfsl-rolling-window}. The window of the samples is run through Algorithm~\ref{alg:traffic_pattern_match} to evaluate the observed behavior against the learned behavior. The process continues on a rolling basis as long as the flow is active.

\noindent \textbf{Model Effectiveness:} From our evaluation results, we observe several key strengths of the proposed model. First, the fine-grained analysis provided by our novel CR graph enables the extraction of transmission behaviors along the graph's edges. Second, this level of analysis allows us to identify not only deviations in attack patterns, such as flooding attacks, but also deviations in between other benign traffic patterns, enabling the model to distinguish between different uses of the granted network permissions. Finally, we found that the approach demonstrates a reasonable resilience to statistical fluctuations caused by changing network conditions, including delays and jitter. Notably, the model remains effective even when trained under conditions different from those present during enforcement, showcasing its adaptability across diverse operational scenarios.

%% file: rgam_module.tex
\section{The Rule Generator and Association Miner Module (RGAM)}
\label{sec:rgam}

This module has a twofold objective. 
First, it generates network access control rules specific to each edge in the CR graph based on the observed packet traffic during training. 
Our approach automatically infers which protocol header attributes (i.e., predicates) from the observed traffic should be used to construct those rules. For instance, the predicates could be the ethernet type, ARP operation type, ICMP code, protocol codes, IP addressing and transport layer ports. In the case of the example in Figure~\ref{fig:com-req}, it will identify which predicates are needed for permitting each of the four flows. The generated rules will eventually be installed in the network switches if the appropriate access request is made. 

This module, therefore, generates only \textit{possitive rules}, meaning that the generated rules only allow the appropriate traffic to pass. Traffic that does not match any generated rules will be matched to the default rule and, therefore, sent to the controller. The ARL makes the access control decision and denies the traffic if it doesn't belong to the learned packet distribution.

Second, the RGAM utilizes the generated rules to identify strong associations among them. The underlying concept is that when a rule is scheduled for deployment based on an access request, it is efficient to deploy all strongly associated rules proactively. Such a strategy helps minimize the exchange of \verb|OFPT_PACKET_IN|, \verb|OFPT_PACKET_OUT|, and \verb|OFPT_FLOW_MOD| messages between the control and network switches, thereby reducing the overhead caused by these control plane operations. Moreover, identifying rule associations enables ZT-SDN to learn the unique communication characteristics of different protocols. For example, in the case of TCP protocol, bidirectional flows are required between the endpoints, while in UDP, the receiver is not obligated to respond to the sender. In the illustrated example in Figure~\ref{fig:com-req}, RGAM finds strong correlations between rules of the same protocol stack (i.e., rule pairs for flows \circled{1}--\circled{2}, and \circled{3}--\circled{4} are strongly associated) as well as between rules of different protocol stacks (i.e., rules for flows \circled{1} to \circled{4} are strongly associated). Therefore, once the streaming application makes an approved access request for the authentication server, all the permitted flows are proactively deployed, saving communication overhead and allowing the application to operate successfully.

The RGAM is based on association rule mining, a technique for discovering relationships and patterns in datasets~\cite{kumbhare2014overview, agarwal1994fast}. It focuses on identifying frequent itemsets and extracting association rules based on co-occurrence patterns in the data. Association rules capture the dependencies and associations between different items or attributes in a dataset. These rules are typically in the form of $\{antecedents\} \Rightarrow \{consequences\}$ statements, where certain attributes in a transaction (i.e., the antecedents) imply the presence of other attributes (i.e., the consequences). 

Each generated rule has metrics to measure how strong the association is. First, the \textit{support} metric is defined as
\begin{equation}
    P(A \cap B) = \frac{freq(A,B)}{N}    
\end{equation}
where $A$ and $B$ are itemsets (i.e., sets of one or more attributes) of various lengths, and $P(A \cap B)$ is the joined probability of $A$ and $B$. Thus, support measures the occurrence of a rule in a dataset. Second, the \textit{confidence} metric is the conditional probability defined as 
\begin{equation}
    P(B|A)=\frac{support(A, B)}{support(A)}    
\end{equation}
Thus, confidence measures how often itemset $B$ appears with the itemset $A$ given the frequency of $A$.

\noindent\textbf{Rule Generator.} It takes as input the CR graph and an application represented as a node in the CR graph. The output is the inferred rules. A rule is defined as
\begin{equation}
    rule=[key_1=value_1,\ldots, key_n=value_n]   
\end{equation}
where a $key$ is a protocol header name (e.g., $source\_ip$), $value$ is the value of that header, and $n$ is the rule's cardinality. A network packet matches a rule if all the header-value pairs present in the rule are also present in the packet's header.

Algorithm~\ref{alg:rule_generator} shows the pseudo-code of our rule generator algorithm. We limit the processing to all the packet header features available for enforcement per the OpenFlow standard~\cite{openflowSwitch}, such as VLAN ID, Ethernet/IP addresses, protocol flags and options, and ports.
First, we fetch a subgraph (i.e., a part of the CR graph), which contains all the nodes and edges (i.e., streams) for which the application is a source or a sink (that is, the application and the destination services). 
The traffic pertaining to the edges of the CR subgraph is essentially the traffic datasets used to train the ARL module. As mentioned earlier, the ARL determines those datasets' length dynamically until the distribution is learned. Once the dataset sufficiently captures the distribution, it is used to mine access control rules, a one-time process.

Subsequently, the algorithm prepares the binary tables, one of each stream direction (lines~\ref{alg:rule_generator:columns_start}-\ref{alg:rule_generator:columns_end}). The columns of a binary table are the unique header-value pairs of each column in the flow dataset. For instance, if the $source\_port$ attribute has three unique values in the dataset, 4455, 5588, and 6699, then that would result in three columns in the binary table: $source\_port:4455$, $source\_port:5588$, and $source\_port:6699$. The number of rows would equal the total number of packets in the stream. 

Afterward, the algorithm populates the binary tables. For each packet (i.e., row), we set the value of a cell to 1 if the condition of the column holds for the packet; otherwise, we set it to 0. Finally, we apply the Apriori algorithm~\cite{singh2013improving} (line~\ref{alg:rule_generator:apriori}) with a pre-determined minimum support value (90\% for the experiments). The generated frequent itemsets are then used to generate the association rules (line~\ref{alg:rule_generator:assoc_rules}). The algorithm uses a high minimum confidence score threshold (100\% for the experiments) to eliminate the less confident associations. For each generated rule, we then unify the antecedences and consequences (line~\ref{alg:rule_generator:unify}) to create sets of header-value pairs and discard any duplicate sets (line~\ref{alg:rule_generator:unique}). Finally, the algorithm stores the rules of maximum cardinality.

\begin{algorithm}
\caption{Rule Generator}\label{alg:rule_generator}
\begin{algorithmic}[1]
\Require An application identifier, CR graph 
% \State $app\_data \gets drop\_features(app\_data)$ \label{alg:rule_generator:filter_columns}
\State $all\_streams \gets CR.$get\_steams($app$)\label{alg:rule_generator:decompose}
\State $binary\_tables \gets$ \O
\For {$stream$ in $all\_streams$} \label{alg:rule_generator:columns_start}
\State $columns \gets$ \O
\For {$column$ in $stream.columns$}
\State $columns \gets columns +$ get\_unique($column$)
\EndFor
\State $table \gets steam.no\_packets \times columns$
\State $binary\_tables.$add($table$)
\EndFor \label{alg:rule_generator:columns_end}
\State populate($binary\_tables$)
\State $generated\_rules \gets$ \O
\For {$table$ in $binary\_tables$}
\State $freq\_items \gets$ apriori($table, min\_support$) \label{alg:rule_generator:apriori}
\State $rules \gets$ assoc\_rules($freq\_items, min\_confidence$)\label{alg:rule_generator:assoc_rules} 
\State $rules \gets rules.antecedents \cup rules.consequents$\label{alg:rule_generator:unify}
\State $rules \gets$ unique($rules$)\label{alg:rule_generator:unique}
\State $generated\_rules.$add(max\_cardinality($rules$)) 
\EndFor
\end{algorithmic}
\end{algorithm}

The generated rules are considered \textit{correct} due to the use of high support and confidence scores. However, \textit{completeness} is not always guaranteed, as this requirement heavily relies on the intricacies of applications. For instance, consider an application that employs two communication channels with a particular service. If the second channel rarely appears in the dataset, there might be scenarios where a rule allowing this infrequent flow is not generated. This is because communication occurs so rarely that no rule with sufficiently high confidence or support can be derived. To address this issue, ZT-SDN generates such rules at access request time. This means that if a \verb|PACKET_IN| request is authorized by the ARL module, and the generated rules for the application have already been installed, but a packet does not match any of those rules, ZT-SDN dynamically creates a proprietary rule based on the packet's source/destination address, port numbers, and protocol. This ensures that the transmission can be accommodated even for flows that were infrequent or not explicitly extracted during the rule generation phase.

\noindent\textbf{Rule Association Miner.} Once the rule generator computes the rules, we find strong associations between those rules (see Algorithm~\ref{alg:rule_assoc_miner}). 
First, we construct a binary table (line~\ref{alg:rule_assoc_miner:table}) with columns representing the generated rules from Algorithm~\ref{alg:rule_generator}. Then, we compute the initial time window (lines~\ref{alg:rule_assoc_miner:win_start} to \ref{alg:rule_assoc_miner:win_end}) and get the packets within this window (line~\ref{alg:rule_assoc_miner:packets}). The algorithm checks which rules (i.e., columns) apply to the packets in the window and sets the corresponding cells to 1; otherwise, to 0 (line~\ref{alg:rule_assoc_miner:populate}). Then, the next time window is computed (lines~\ref{alg:rule_assoc_miner:next_win_start} to \ref{alg:rule_assoc_miner:next_win_end}). We use packet indexes to compute the time window, as the application dataset may have time gaps. Thus, the algorithm saves a lot of loop iterations by skipping time windows that do not include any packets. Finally, the rule association is computed using the apriori algorithm (lines~\ref{alg:rule_assoc_miner:apriori} to \ref{alg:rule_assoc_miner:assoc_rules}) while setting the minimum support and confidence to high values (90\% and 100\% for the experiments respectively).

\begin{algorithm}
\caption{Rule Association Miner}\label{alg:rule_assoc_miner}
\begin{algorithmic}[1]
\Require Generated rules, Application's traffic dataset
\State $binary\_table \gets 1 \times rules$\label{alg:rule_assoc_miner:table}
\While{True}
\If{$ws = 0$}\label{alg:rule_assoc_miner:win_start}
\State $packet\_time \gets app\_data[0].timestamp$
\State $ws \gets packet\_time$
\EndIf
\State $we \gets ws + w\_duration$\label{alg:rule_assoc_miner:win_end}
\State $packets \gets$ get\_packets($app\_data, [ws, we]$)\label{alg:rule_assoc_miner:packets}
\State populate($binary\_table$)\label{alg:rule_assoc_miner:populate} 
\State $index \gets$ last\_packet($packets$)$.index$\label{alg:rule_assoc_miner:next_win_start} 
\If{$index + 1 < app\_data.length$}
\State $index \gets index + 1$
\State $ws \gets app\_data[index].timestamp$
\Else ~~$break$
% \State $break$
\EndIf\label{alg:rule_assoc_miner:next_win_end} 
\EndWhile
\State $freq\_items \gets$ apriori($binary\_table, min\_support$) \label{alg:rule_assoc_miner:apriori}
\State $associations \gets$ assoc\_rules($freq\_items,$\\
\hspace{\algorithmicindent}$min\_confidence$)\label{alg:rule_assoc_miner:assoc_rules} 
\end{algorithmic}%
\end{algorithm}

%% file: zt-gym.tex
\section{ZT-Gym}
\label{sec:zt-gym}

In this section, we present our offline approach for creating application network datasets. The datasets are generated within a controlled environment devoid of any adversarial presence, ensuring the production of benign data. These generated datasets are then employed in the ZT-SDN training mode as outlined in Section~\ref{sec:arch-overview}.

\subsection{The ZT-Gym Environment}

\begin{figure}[htbp]
\centerline{\includegraphics[width=.6\columnwidth]{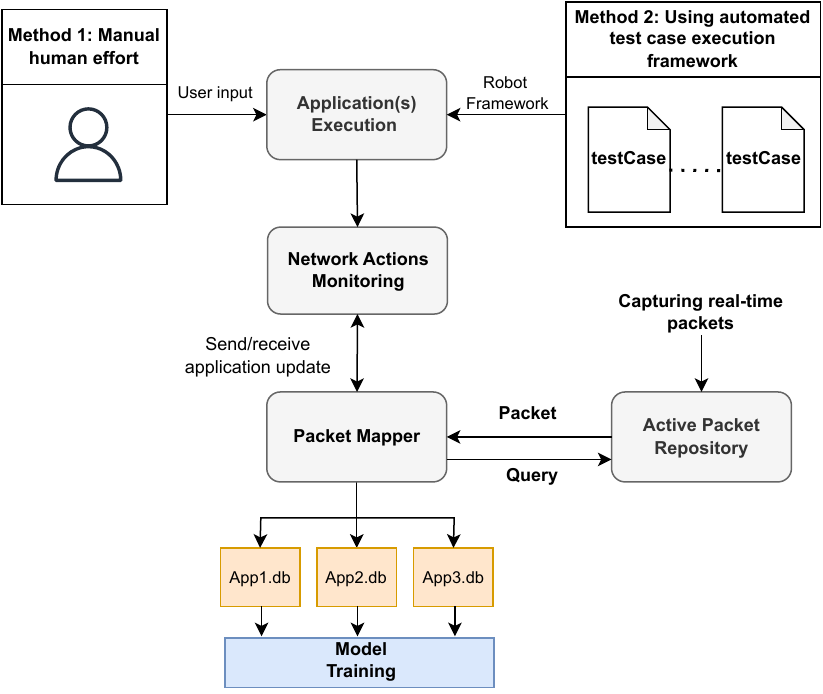}}
\caption{
The ZT-Gym environment.
\label{fig:zt-gym}
}
\end{figure}

We developed ZT-Gym as a Linux-based virtual machine and a dataset generation pipeline shown in Figure~\ref{fig:zt-gym}. ZT-Gym supports the dataset generation for several applications simultaneously. The actual applications or services are installed within the ZT-Gym environment. The data (i.e., network traffic) is obtained directly from the network interface(s) to ensure that the generated data are realistic and accurate. 

\subsection{Technical Details}

\noindent{\bf Application input methods.} To generate network data, the application requires input. We support two input methods: (1) Manual human interaction – a human user interacts with the application in the virtual environment, performing the tasks it would typically do in normal conditions, and (2) an existing framework for application testing automation.

The first method is straightforward but requires human time, which is expensive. To address this, we use the Robot Framework~\cite{robotf, robot-boon}, an open-source automation framework that streamlines the execution flows of applications with graphical user interfaces or those run via the command line. Constructing test cases using the Robot Framework is an intuitive and straightforward process. To write a Robot Framework test case, one needs to define settings, variables, and test cases in a plain-text file with the ``.robot'' extension. The Robot Framework is widely adopted in corporate settings for automated testing. Companies often possess their own .robot files tailored for applications, developed as part of the software development life cycle. In this method, the test case executes the actions that the application would normally exhibit in benign conditions.

\noindent{\bf Active packet repository.} It captures real-time packets exchanged within the ZT-Gym environment. The packets are captured using the ``tcpdump'' tool and are stored in a MongoDB database. These packets are stored in real-time, and the database is updated constantly to accommodate any new packets generated or received by the system.

\noindent{\bf Network Actions Monitoring.} Implemented as a Python script with administrative privileges within the ZT-Gym environment, this module monitors network actions specific to the executing application(s). It focuses on aspects such as port bindings and the transport layer protocols employed for transmission through these ports. Notably, this module replicates the functionality described in the \textit{Host Module} (Section~\ref{sec:hm}).
To achieve its monitoring objectives, the script uses the Linux utility \verb|ss -plan| to identify the ports bound by the applications~\cite{ss}. Subsequently, it transmits a tuple to the packet mapper, encompassing information such as the application name, port numbers in use, destination IP address being communicated with, and the timestamp. Any updates in these features trigger notifications to the Packet Mapper module, facilitating the classification of packets to their respective applications.

\noindent{\bf Packet Mapper.} It plays a crucial role in associating network packets with specific applications. This module resembles part of the functionality of the \textit{Controller-Specific Module} described in Section~\ref{sec:csm}
In this mapping process, five key features are considered: the packet's timestamp, source and destination IP addresses, and source and destination ports. Timestamp is a pivotal parameter for the mapping packets to specific applications running in ZT-Gym. This is because different applications may use the same port(s) at different points in time. 
To achieve this, the module leverages the timestamp provided by the Network Actions Monitoring module. Packets transmitted from this timestamp onward are attributed to the corresponding application identified in the tuple. In cases where the module detects another application utilizing the same port at a later time, packets sent or received up until the second application binds the port are categorized under the purview of the first application. The approach ensures accurate packet classification.

\noindent{\bf Model Training.} This module takes the generated datasets, containing benign network data from applications, as input. Its functionality for training the models mirrors that of the \textit{ARL} (Section~\ref{sec:arl}), \textit{RTFSL} (Section~\ref{sec:rtfsl}) and \textit{RGAM} (Section~\ref{sec:rgam}) modules, including the model retraining until the user-specified heuristics are met. 

\noindent\textbf{Fidelity.} Note that every component within the ZT-Gym environment replicates the exact functionality of its counterpart in the ZT-SDN architecture (Figure~\ref{fig:architecture}). The key distinction lies in the ZT-Gym environment's controlled setting, where data is generated. In this environment, the ML models are trained offline using the generated datasets. In contrast, ZT-SDN learns directly from data generated and exchanged in the underlying network. 
Since the applications scrutinized are the same as the ones expected to run in the network, the corresponding packet-level features (e.g., payload) are identical. Some packet-level features, such as TTL, are standardized by the operating systems. However, flow statistics may vary from the actual network due to different conditions, such as lower bandwidth. As demonstrated in Sections \ref{sec:rtfsl-effectiveness} and \ref{sec:rtfsl-reduced-bandwidth}, RTFSL can tolerate reasonable degradations. Administrators can also adjust the bandwidth within the Gym environment~\cite{tcLinux}, allowing for data generation that closely mirrors their networks.

%% file: evaluation.tex
\section{Evaluation}

We focus on answering the following research questions (RQ):

\begin{itemize}
    \item \textbf{RQ1:} How effective is the ARL module in enforcing network access control?

    \item \textbf{RQ2:} Why is KitNet the preferred learning model for the ARL module, and how do baseline models compare in terms of performance?
    
    \item \textbf{RQ3:} Once access has been granted, how effective is the RTFSL module in ensuring that the usage of the deployed policies adheres to the leaned benign behavior? 
    
    \item \textbf{RQ4:} How do network delays impact the RTFSL model's effectiveness?

    \item \textbf{RQ5:} Is ZT-SDN scalable when applied to an SDN network? Does it degrade the network throughput?
    
\end{itemize}

\subsection{RQ1: ARL Module Effectiveness}
\label{sec:eval-ac}

The access control aspect concerns the evaluation of our module's ability to correctly allow or deny network accesses during the ML module enforcement. 

\noindent\textbf{Experiment Setup.} 
We evaluated the effectiveness of the ARL module using two experimental setups: (1) a real-world traffic dataset captured at a major transit link, and (2) synthetic traffic generated within Mininet, a software-defined networking (SDN) environment.

\underline{Setup 1 -- Real traffic dataset:} 
For the real-world traffic evaluation, we used the MAWI traffic datasets~\cite{mawi_wide_dataset}, a comprehensive collection of Internet traffic data from the WIDE (Widely Integrated Distributed Environment) project. The dataset, monitored by the MAWI Working Group since 2001, includes packet traces from trans-Pacific links between Japan and the United States.

We selected network traffic data from two periods: 2018 (older data) and 2024 (recent data). For each period, we analyzed traffic from prominent protocol stacks, specifically \verb|ETHERNET_IP_TCP| and \verb|ETHERNET_IP_UDP|. Services were chosen based on the following criteria: (1) At least one client user or application connecting to the service. (2) A sufficient volume of packet data for model training and evaluation (minimum of 20,000 packets).
(3) Adequate flow duration to capture transmission behavior (minimum of 10 minutes).

While the MAWI dataset lacks direct information about the applications generating the traffic, we inferred application types by mapping port numbers to services using the SpeedGuide database~\cite{speedguide}. SpeedGuide is a comprehensive online resource offering an extensive database of TCP and UDP port numbers. It provides details about well-known, registered, and private port assignments, along with information on associated protocols and services. Table~\ref{tab:services} summarizes the selected services, their port numbers, and protocol stacks. We extracted data from four service categories: HTTP, HTTPS, Gaming/Web Calling, and Voice over IP (VoIP).

\begin{table}[]
\caption{Services used for evaluation from the MAWI dataset.}
\label{tab:services}
\resizebox{\textwidth}{!}{%
\begin{tabular}{@{}|c|c|c|l|@{}}
\toprule
\textbf{Service Port Number} & \textbf{Service} & \textbf{Protocol} & \multicolumn{1}{c|}{\textbf{Description}}                                          \\ \midrule
80                           & HTTP             & TCP               & Hyper Text Transfer Protocol (HTTP) - port used for web traffic.                   \\ \midrule
443                          & HTTPS            & TCP               & HTTPS / SSL - encrypted web traffic, also used for VPN tunnels over HTTP.          \\ \midrule
3479 &
  Games &
  UDP &
  \begin{tabular}[c]{@{}l@{}}Microsoft Teams uses UDP ports 3478 through 3481 for media traffic, as well as TCP ports 80 and 443. \\ Apple FaceTime, Apple Game Center use ports 3478-3497 (UDP). \\ Call of Duty World at War.\\ Playstation 4 game ports.\end{tabular} \\ \midrule
19305                        & VoIP             & UDP               & Google Talk, DUO, Hangouts commonly use ports 19302-19308 UDP and 19305-19308 TCP. \\ \bottomrule
\end{tabular}%
}
\end{table}

\underline{Setup 2 -- Using Mininet:}
We utilized Mininet to simulate a controlled network environment. A Linux-based server was deployed in one container, while a network client was set up in another container, both connected to the same virtual network.
The experiments revolved around generating realistic network traffic between these two containers, for which we employed the \verb|iperf| tool. The client uses two communication channels based on the \verb|ETHERNET_IP_TCP| and another two channels based on the \verb|ETHERNET_IP_UDP| protocol stacks to communicate with the server. Those channels are initiated from a random source port in every session on the client side and directed to ports 5001 (TCP) and 5010 (UDP) on the server side. The application transmits data as fast as possible, leveraging as much of the available bandwidth as possible.

\noindent \textbf{Model Training.} 
For this module, we trained the model with the following parameter settings for all applications: feature mapping samples were set to 200~\cite{mirsky2018kitsune}, and the minimum number of training samples was 20K, while the maximum reconstruction error (computed as Root-Mean-Square Error -- RMSE) training and validation periods were limited to 0.009 and 0.05 respectively.

\noindent \textbf{Model Testing.} 
We evaluate the efficacy of the ARL module in three distinct settings:

\begin{enumerate}
    \item \textit{Normal Benign Performance:} This setting assesses the modules's ability to perform on unseen (future) benign data from the same application running on the same host machine. 

    \item \textit{Abnormal Benign Detection:} In this setting, we test whether the module can identify deviations in benign traffic patterns. Specifically, we send benign traffic packets generated by other applications (either of the same or different type) using the same protocol stack to determine if the model flags them as anomalous. Note that we change the source and destination IP to the ones on which the model was trained. Thus, the module treats the packets as packets sent from the same application on which it has been trained. These deviations are categorized as ``abnormal benign'' because, while the traffic originates from benign applications, it diverges from the patterns learned by the model.

    \item \textit{Abnormal Malicious Detection:} To evaluate the module's ability to detect malicious traffic, we use two widely recognized attacks: flooding and port scanning. These attacks were chosen for their prevalence in testing anomaly detection systems~\cite{giotis2014combining, nguyen2008network, zavrak2020anomaly}, their prominence as network threats, and their potential to induce significant behavioral deviations in communication flows monitored by ZT-SDN. The attacks were executed using the Linux tool \verb|hping3|.
\end{enumerate}

\begin{table}[]
\caption{ARL evaluation results for UDP-based services. The ``e'' notation denotes a power of 10.}
\label{tab:udp-arl-results}
\resizebox{1.1\textwidth}{!}{%
\begin{tabular}{|cl|ll|ll|ll|ll|ll|ll|ll|ll|}
\hline
\multicolumn{2}{|c|}{\textbf{\begin{tabular}[c]{@{}c@{}}Testing (Unseen) $\rightarrow$\\ \\ Training $\downarrow$\end{tabular}}} &
  \multicolumn{2}{c|}{\textbf{2018 VoIP 1 (UDP)}} &
  \multicolumn{2}{c|}{\textbf{2018 VoIP 2 (UDP)}} &
  \multicolumn{2}{c|}{\textbf{2019 VoIP 3 (UDP)}} &
  \multicolumn{2}{c|}{\textbf{2024 Game 1 (UDP)}} &
  \multicolumn{2}{c|}{\textbf{2018 VoIP 4 (UDP)}} &
  \multicolumn{2}{c|}{\textbf{iperf (udp)}} &
  \multicolumn{2}{c|}{\textbf{Flooding (UDP)}} &
  \multicolumn{2}{c|}{\textbf{PortScan (UDP)}} \\ \hline
\multicolumn{2}{|c|}{\textbf{\begin{tabular}[c]{@{}c@{}}2018 VoIP 1 (UDP)\\ \\ Training stop: 20204\\ Max validation score: 0.023\end{tabular}}} &
  \multicolumn{2}{l|}{\cellcolor[HTML]{FFCE93}\begin{tabular}[c]{@{}l@{}}Total: 203409\\ TP: 0\\ TN: 203409\\ FP: 0\\ FN: 0\\ Max Error: 0.023\\ Min Error: 4.28e-5\end{tabular}} &
  \multicolumn{2}{l|}{\cellcolor[HTML]{9AFF99}\begin{tabular}[c]{@{}l@{}}Total: 1449\\ TP: 1449\\ TN: 0\\ FP: 0\\ FN: 0\\ Max Error: 2.57e18\\ Min Error: 2.57e18\end{tabular}} &
  \multicolumn{2}{l|}{\cellcolor[HTML]{9AFF99}\begin{tabular}[c]{@{}l@{}}Total: 1424\\ TP: 1424\\  TN: 0\\      FP: 0\\      FN: 0\\      Max Error: 4.9e18\\      Min Error: 4.9e18\end{tabular}} &
  \multicolumn{2}{l|}{\cellcolor[HTML]{9AFF99}\begin{tabular}[c]{@{}l@{}}Total: 32872\\      TP: 32872\\      TN: 0\\      FP: 0\\      FN: 0\\      Max Error: 8.67e19\\      Min Error: 8.67e19\end{tabular}} &
  \multicolumn{2}{l|}{\cellcolor[HTML]{9AFF99}\begin{tabular}[c]{@{}l@{}}Total: 335221\\      TP: 335221\\      TN: 0\\      FP: 0\\      FN: 0\\      Max Error: 2.09e19\\      Min Error: 2.09e19\end{tabular}} &
  \multicolumn{2}{l|}{\cellcolor[HTML]{9AFF99}\begin{tabular}[c]{@{}l@{}}Total: 35672\\      TP: 35672\\      TN: 0\\      FP: 0\\      FN: 0\\      Max Error: 1.22e20\\      Min Error: 7.21e19\end{tabular}} &
  \multicolumn{2}{l|}{\cellcolor[HTML]{FFCCC9}\begin{tabular}[c]{@{}l@{}}Total: 24753\\      TP: 24753\\      TN: 0\\      FP: 0\\      FN: 0\\      Max Error: 2.6e20\\      Min Error: 7.2e19\end{tabular}} &
  \multicolumn{2}{l|}{\cellcolor[HTML]{FFCCC9}\begin{tabular}[c]{@{}l@{}}Total: 5282\\      TP: 5282\\      TN: 0\\      FP: 0\\      FN: 0\\      Max Error: 2.81e20\\      Min Error: 2.80e20\end{tabular}} \\ \hline
\multicolumn{2}{|c|}{\textbf{\begin{tabular}[c]{@{}c@{}}2024 Game 1 (UDP)\\ \\ Training stop: 20200 \\ Max validation score: 0.006\end{tabular}}} &
  \multicolumn{2}{l|}{\cellcolor[HTML]{9AFF99}\begin{tabular}[c]{@{}l@{}}Total: 170646\\      TP: 170646\\      TN: 0 \\      FP: 0\\      FN: 0\\      Max Error: 8.677e19\\      Min Error: 8.67e19\end{tabular}} &
  \multicolumn{2}{l|}{\cellcolor[HTML]{9AFF99}\begin{tabular}[c]{@{}l@{}}Total: 1449\\      TP: 1449\\      TN: 0\\      FP: 0\\      FN: 0\\      Max Error: 8.57e19\\      Min Error: 8.57e19\end{tabular}} &
  \multicolumn{2}{l|}{\cellcolor[HTML]{9AFF99}\begin{tabular}[c]{@{}l@{}}Total: 1424\\      TP: 1424\\      TN: 0\\      FP: 0\\      FN: 0\\      Max Error: 8.49e19\\      Min Error: 8.49e19\end{tabular}} &
  \multicolumn{2}{l|}{\cellcolor[HTML]{FFCE93}\begin{tabular}[c]{@{}l@{}}Total: 39899\\      TP: 0\\      TN: 39899\\      FP: 0\\      FN: 0\\      Max Error: .006\\      Min Error: 4.05e-05\end{tabular}} &
  \multicolumn{2}{l|}{\cellcolor[HTML]{9AFF99}\begin{tabular}[c]{@{}l@{}}Total: 335221\\      TP: 335221\\      TN: 0\\      FP: 0\\      FN: 0\\      Max Error: 9.68e19\\      Min Error: 9.68e19\end{tabular}} &
  \multicolumn{2}{l|}{\cellcolor[HTML]{9AFF99}\begin{tabular}[c]{@{}l@{}}Total: 35672\\      TP: 35672\\      TN: 0\\      FP: 0\\      FN: 0\\      Max Error: 6.53e19\\      Min Error: 3.05e19\end{tabular}} &
  \multicolumn{2}{l|}{\cellcolor[HTML]{FFCCC9}\begin{tabular}[c]{@{}l@{}}Total: 24753\\      TP: 24753\\      TN: 0\\      FP: 0\\      FN: 0\\      Max Error: 2.16e20\\      Min Error: 7.73e18\end{tabular}} &
  \multicolumn{2}{l|}{\cellcolor[HTML]{FFCCC9}\begin{tabular}[c]{@{}l@{}}Total: 5282\\      TP: 5282\\      TN: 0\\      FP: 0\\      FN: 0\\      Max Error: 2.38e20\\      Min Error: 2.38e20\end{tabular}} \\ \hline
\multicolumn{2}{|c|}{\textbf{\begin{tabular}[c]{@{}c@{}}2018 VoIP 4 (UDP)\\ \\ Training stop: 20200\\ Max validation score: 0.026\end{tabular}}} &
  \multicolumn{2}{l|}{\cellcolor[HTML]{9AFF99}\begin{tabular}[c]{@{}l@{}}Total: 170646\\      TP: 170646\\      TN: 0 \\      FP: 0\\      FN: 0\\      Max Error: 3.37e17\\      Min Error: 3.37e17\end{tabular}} &
  \multicolumn{2}{l|}{\cellcolor[HTML]{9AFF99}\begin{tabular}[c]{@{}l@{}}Total: 1449\\      TP: 1449\\      TN: 0\\      FP: 0\\      FN: 0\\      Max Error: 2.34e19\\      Min Error: 2.34e19\end{tabular}} &
  \multicolumn{2}{l|}{\cellcolor[HTML]{9AFF99}\begin{tabular}[c]{@{}l@{}}Total: 1424\\      TP: 1424\\      TN: 0\\      FP: 0\\      FN: 0\\      Max Error: 2.58e19\\      Min Error: 2.58e19\end{tabular}} &
  \multicolumn{2}{l|}{\cellcolor[HTML]{9AFF99}\begin{tabular}[c]{@{}l@{}}Total: 32872\\      TP: 32872\\      TN: 0\\      FP: 0\\      FN: 0\\      Max Error: 9.68e19\\      Min Error: 9.68e19\end{tabular}} &
  \multicolumn{2}{l|}{\cellcolor[HTML]{FFCE93}\begin{tabular}[c]{@{}l@{}}Total: 386575\\      TP: 0\\      TN: 386575\\      FP: 0\\      FN: 0\\      Max Error: 0.026\\      Min Error: 4.28e-5\end{tabular}} &
  \multicolumn{2}{l|}{\cellcolor[HTML]{9AFF99}\begin{tabular}[c]{@{}l@{}}Total: 35672\\      TP: 35672\\      TN: 0\\      FP: 0\\      FN: 0\\      Max Error: 1.4e20\\      Min Error: 7.28e19\end{tabular}} &
  \multicolumn{2}{l|}{\cellcolor[HTML]{FFCCC9}\begin{tabular}[c]{@{}l@{}}Total: 24753\\      TP: 24753\\      TN: 0\\      FP: 0\\      FN: 0\\      Max Error: 7.2e19\\      Min Error: 2.8e19\end{tabular}} &
  \multicolumn{2}{l|}{\cellcolor[HTML]{FFCCC9}\begin{tabular}[c]{@{}l@{}}Total: 5282\\      TP: 5282\\      TN: 0\\      FP: 0\\      FN: 0\\      Max Error: 3.01e20\\      Min Error: 3e20\end{tabular}} \\ \hline
\multicolumn{2}{|c|}{\textbf{\begin{tabular}[c]{@{}c@{}}iperf (UDP)\\      \\ Training stop: 20201\\      Max validation score: 0.021\end{tabular}}} &
  \multicolumn{2}{l|}{\cellcolor[HTML]{9AFF99}\begin{tabular}[c]{@{}l@{}}Total: 170646\\      TP: 170646\\      TN: 0 \\      FP: 0\\      FN: 0\\      Max Error: 7.11e19\\      Min Error: 7.04e19\end{tabular}} &
  \multicolumn{2}{l|}{\cellcolor[HTML]{9AFF99}\begin{tabular}[c]{@{}l@{}}Total: 1449\\      TP: 1449\\      TN: 0\\      FP: 0\\      FN: 0\\      Max Error: 7.11e19\\      Min Error: 7.10e19\end{tabular}} &
  \multicolumn{2}{l|}{\cellcolor[HTML]{9AFF99}\begin{tabular}[c]{@{}l@{}}Total: 1424\\      TP: 1424\\      TN: 0\\      FP: 0\\      FN: 0\\      Max Error: 7.11e19\\      Min Error: 7.1e19\end{tabular}} &
  \multicolumn{2}{l|}{\cellcolor[HTML]{9AFF99}\begin{tabular}[c]{@{}l@{}}Total: 32872\\      TP: 32872\\      TN: 0\\      FP: 0\\      FN: 0\\      Max Error: 1.24e19\\      Min Error: 7.72e18\end{tabular}} &
  \multicolumn{2}{l|}{\cellcolor[HTML]{9AFF99}\begin{tabular}[c]{@{}l@{}}Total: 335221\\      TP: 335221\\      TN: 0\\      FP: 0\\      FN: 0\\      Max Error: 7.11e19\\      Min Error: 7.04e19\end{tabular}} &
  \multicolumn{2}{l|}{\cellcolor[HTML]{FFCE93}\begin{tabular}[c]{@{}l@{}}Total: 304952\\      TP: 0\\      TN: 304952\\      FP: 0\\      FN: 0\\      Max Error: 0.07\\      Min Error: 0.0018\end{tabular}} &
  \multicolumn{2}{l|}{\cellcolor[HTML]{FFCCC9}\begin{tabular}[c]{@{}l@{}}Total: 24753\\      TP: 24753\\      TN: 0\\      FP: 0\\      FN: 0\\      Max Error: 1.02e19\\      Min Error: 1.02e19\end{tabular}} &
  \multicolumn{2}{l|}{\cellcolor[HTML]{FFCCC9}\begin{tabular}[c]{@{}l@{}}Total: 5282\\      TP: 5282\\      TN: 0\\      FP: 0\\      FN: 0\\      Max Error: 1.54e20\\      Min Error: 7.18e18\end{tabular}} \\ \hline
\end{tabular}%
}
\end{table}

\noindent \textbf{Results.} 
We assess the detection performance of this module for the different attack and benign data, employing the following metrics: true positives (TP), true negatives (TN), false positives (FP), and false negatives (FN). TP and FN are relevant to the abnormal benign or malicious data, indicating the proportion of examples correctly classified as deviant or mistakenly classified as normal benign. On the other hand, TN and FP pertain to normal benign data, representing the portion of examples correctly classified as normal benign or mistakenly classified as deviant. 

Table~\ref{tab:udp-arl-results} shows the ARL prediction results for the UDP-based applications/services. The table presents the results in a cross-evaluation manner where the vertical axis indicates the data on which the model is individually trained. For each training data, we report the number of training samples supplied to the model as well as the maximum RMSE reported during the validation period. The horizontal axis indicates the data on which the model is tested (unseen traffic data). The orange cells denote the test to evaluate the effectiveness of the model in recognizing normal benign data, whereas the green cells evaluate the effectiveness of the model in recognizing abnormal benign data. Finally, the red cells indicate the effectiveness of the model in recognizing abnormal malicious data. In each cell, we report the total number of packets on which the model was tested as well as the TP, TN, FP, and FN numbers. We also report the minimum and maximum RMSE errors reported by the model while testing in those packets. The column/row naming follows the following convention: ``[Dataset Year] [Application Type] [Client ID] ([Protocol]).'' Attack traffic data and iperf do not have a year as they were generated using tools. Note that the ``e'' notation denotes a power of 10. 

Our findings demonstrate that the module achieves a 100\% accuracy in identifying normal benign traffic, effectively learning the distribution for each of the CR graph edges with just 20,000 training samples. Furthermore, the model exhibits a 100\% accuracy in detecting anomalous benign packets, showcasing its ability to discern differences in protocol header values even when client applications connect to the same service (e.g., VoIP-to-VoIP communication). Last, the model maintains 100\% accuracy in detecting anomalous malicious traffic. In these scenarios, the port number is a key feature, as malicious activities like flooding and port scanning rely on a range of source or destination ports not utilized by the benign applications on which the model is trained.

Tables~\ref{tab:tcp-arl-results-1} and \ref{tab:tcp-arl-results-2} show the ARL performance results for TCP-based applications/services. As the table shows, the ARL training requires more packet samples from each application as the TCP protocol stack contains more features compared to the UDP stack. Thus, more training data is required for our unsupervised model to understand the relation between header-value pairs. The results are consistent with the ones presented in Table~\ref{tab:udp-arl-results}. The ARL module can distinguish between normal benign and abnormal malicious or abnormal benign accesses.

\begin{table}[]
\caption{ARL evaluation results for TCP-based services.}
\label{tab:tcp-arl-results-1}
\resizebox{\textwidth}{!}{%
\begin{tabular}{@{}|cl|ll|ll|ll|ll|ll|ll|@{}}
\toprule
\multicolumn{2}{|c|}{\textbf{\begin{tabular}[c]{@{}c@{}}Testing (Unseen) $\rightarrow$\\ \\ Training $\downarrow$\end{tabular}}} &
  \multicolumn{2}{c|}{\textbf{2018 HTTP 1 (TCP)}} &
  \multicolumn{2}{c|}{\textbf{2018 HTTP 3 (TCP)}} &
  \multicolumn{2}{c|}{\textbf{2018 HTTPS 1 (TCP)}} &
  \multicolumn{2}{c|}{\textbf{2018 HTTPS 2 (TCP)}} &
  \multicolumn{2}{c|}{\textbf{2024 HTTP 1 (TCP)}} &
  \multicolumn{2}{c|}{\textbf{2024 HTTP 2 (TCP)}} \\ \midrule
\multicolumn{2}{|c|}{\textbf{\begin{tabular}[c]{@{}c@{}}2018 HTTP 1 (TCP)\\ \\ Training stop: 30200\\ Max validation score: 0.03\end{tabular}}} &
  \multicolumn{2}{l|}{\cellcolor[HTML]{FFCE93}\begin{tabular}[c]{@{}l@{}}Total: 200000\\      TP: 0\\      TN: 200000\\      FP: 0\\      FN: 0\\      Max Error: 0.39\\      Min Error: 0.60\end{tabular}} &
  \multicolumn{2}{l|}{\cellcolor[HTML]{9AFF99}{\color[HTML]{333333} \begin{tabular}[c]{@{}l@{}}Total: 200000\\      TP: 200000\\      TN: \\      FP: 0\\      FN: 0\\      Max Error: 1.04e16 \\      Min Error: 8.4e15\end{tabular}}} &
  \multicolumn{2}{l|}{\cellcolor[HTML]{9AFF99}\begin{tabular}[c]{@{}l@{}}Total: 80796\\      TP: 80796\\      TN: \\      FP: 0\\      FN: 0\\      Max Error: 1.56e18 \\      Min Error: 1.56e18\end{tabular}} &
  \multicolumn{2}{l|}{\cellcolor[HTML]{9AFF99}\begin{tabular}[c]{@{}l@{}}Total: 26107\\      TP: 26107\\      TN: 0\\      FP: 0\\      FN: 0\\      Max Error:  1.56e18\\      Min Error: 1.56e18\end{tabular}} &
  \multicolumn{2}{l|}{\cellcolor[HTML]{9AFF99}\begin{tabular}[c]{@{}l@{}}Total: 134195\\      TP: 134195\\      TN: \\      FP: 0\\      FN: 0\\      Max Error: 2.20e16  \\      Min Error: 2.12e16\end{tabular}} &
  \multicolumn{2}{l|}{\cellcolor[HTML]{9AFF99}\begin{tabular}[c]{@{}l@{}}Total: 22984\\      TP: 22984\\      TN: 0\\      FP: 0\\      FN: 0\\      Max Error: 2.59e17  \\      Min Error: 2.59e17\end{tabular}} \\ \midrule
\multicolumn{2}{|c|}{\textbf{\begin{tabular}[c]{@{}c@{}}2018 HTTPS 1 (TCP)\\ \\ Training stop: 30200\\ Max validation score: 0.13\end{tabular}}} &
  \multicolumn{2}{l|}{\cellcolor[HTML]{9AFF99}\begin{tabular}[c]{@{}l@{}}Total: 199999\\      TP: 199999\\      TN: 0\\      FP: 0\\      FN: 0\\      Max Error: 7.32e19\\      Min Error: 5.51e19\end{tabular}} &
  \multicolumn{2}{l|}{\cellcolor[HTML]{9AFF99}\begin{tabular}[c]{@{}l@{}}Total: 485192\\      TP: 485192\\      TN: 0\\      FP: 0\\      FN: 0\\      Max Error: 2.41e20\\      Min Error: 2.55e18\end{tabular}} &
  \multicolumn{2}{l|}{\cellcolor[HTML]{FFCE93}\begin{tabular}[c]{@{}l@{}}Total: 88528\\      TP: 0\\      TN: 88528\\      FP: 0\\      FN: 0\\      Max Error:  0.13\\      Min Error: 0.004\end{tabular}} &
  \multicolumn{2}{l|}{\cellcolor[HTML]{9AFF99}\begin{tabular}[c]{@{}l@{}}Total: 26107\\      TP: 26107\\      TN: 0\\      FP: 0\\      FN: 0\\      Max Error:  3.66e19\\      Min Error: 3.65e19\end{tabular}} &
  \multicolumn{2}{l|}{\cellcolor[HTML]{9AFF99}\begin{tabular}[c]{@{}l@{}}Total: 134195\\      TP: 134195\\      TN: 0\\      FP: 0\\      FN: 0\\      Max Error: 8.21e19  \\      Min Error: 8.78e18\end{tabular}} &
  \multicolumn{2}{l|}{\cellcolor[HTML]{9AFF99}\begin{tabular}[c]{@{}l@{}}Total: 22984\\      TP: 22984\\      TN: 0\\      FP: 0\\      FN: 0\\      Max Error: 3.18e19  \\      Min Error: 3.06e19\end{tabular}} \\ \midrule
\multicolumn{2}{|c|}{\textbf{\begin{tabular}[c]{@{}c@{}}2024 HTTP 1 (TCP)\\ \\ Training stop: 30200\\ Max validation score: 0.15\end{tabular}}} &
  \multicolumn{2}{l|}{\cellcolor[HTML]{9AFF99}\begin{tabular}[c]{@{}l@{}}Total: 199999\\      TP: 199999\\      TN: 0\\      FP: 0\\      FN: 0\\      Max Error: 2.46e16\\      Min Error: 2.46e16\end{tabular}} &
  \multicolumn{2}{l|}{\cellcolor[HTML]{9AFF99}\begin{tabular}[c]{@{}l@{}}Total: 485192\\      TP: 485192\\      TN: 0\\      FP: 0\\      FN: 0\\      Max Error: 2.65e16\\      Min Error: 2.65e16\end{tabular}} &
  \multicolumn{2}{l|}{\cellcolor[HTML]{9AFF99}\begin{tabular}[c]{@{}l@{}}Total: 80796\\      TP: 80796\\      TN: 0\\      FP: 0\\      FN: 0\\      Max Error:  1.81e18\\      Min Error: 1.81e18\end{tabular}} &
  \multicolumn{2}{l|}{\cellcolor[HTML]{9AFF99}\begin{tabular}[c]{@{}l@{}}Total: 26107\\      TP: 26107\\      TN: 0\\      FP: 0\\      FN: 0\\      Max Error:  1.81e18\\      Min Error: 1.81e18\end{tabular}} &
  \multicolumn{2}{l|}{\cellcolor[HTML]{FFCE93}\begin{tabular}[c]{@{}l@{}}Total: 1170\\      TP: 0\\      TN: 1170\\      FP: 0\\      FN: 0\\      Max Error: 0.59  \\      Min Error: 0.21\end{tabular}} &
  \multicolumn{2}{l|}{\cellcolor[HTML]{9AFF99}\begin{tabular}[c]{@{}l@{}}Total: 22984\\      TP: 22984\\      TN: 0\\      FP: 0\\      FN: 0\\      Max Error: 3.25e17  \\      Min Error: 3.25e17\end{tabular}} \\ \midrule
\multicolumn{2}{|c|}{\textbf{\begin{tabular}[c]{@{}c@{}}2024 HTTPS 2 (TCP)\\ \\ Training stop: 100200\\ Max validation score: 0.011\end{tabular}}} &
  \multicolumn{2}{l|}{\cellcolor[HTML]{9AFF99}\begin{tabular}[c]{@{}l@{}}Total: 199999\\      TP: 199999\\      TN: 0\\      FP: 0\\      FN: 0\\      Max Error: 1.59e18\\      Min Error: 1.59e18\end{tabular}} &
  \multicolumn{2}{l|}{\cellcolor[HTML]{9AFF99}\begin{tabular}[c]{@{}l@{}}Total: 485192\\      TP: 485192\\      TN: 0\\      FP: 0\\      FN: 0\\      Max Error: 1.59e18\\      Min Error: 1.59e18\end{tabular}} &
  \multicolumn{2}{l|}{\cellcolor[HTML]{9AFF99}\begin{tabular}[c]{@{}l@{}}Total: 80796\\      TP: 80796\\      TN: 0\\      FP: 0\\      FN: 0\\      Max Error:  8.8e15\\      Min Error: 8.8e15\end{tabular}} &
  \multicolumn{2}{l|}{\cellcolor[HTML]{9AFF99}\begin{tabular}[c]{@{}l@{}}Total: 26107\\      TP: 26107\\      TN: 0\\      FP: 0\\      FN: 0\\      Max Error:  4.3e15\\      Min Error: 2428.37\end{tabular}} &
  \multicolumn{2}{l|}{\cellcolor[HTML]{9AFF99}\begin{tabular}[c]{@{}l@{}}Total: 134195\\      TP: 134195\\      TN: 0\\      FP: 0\\      FN: 0\\      Max Error: 1.59e18\\      Min Error: 1.59e18\end{tabular}} &
  \multicolumn{2}{l|}{\cellcolor[HTML]{9AFF99}\begin{tabular}[c]{@{}l@{}}Total: 22984\\      TP: 22984\\      TN: 0\\      FP: 0\\      FN: 0\\      Max Error: 1.59e18  \\      Min Error: 1.59e18\end{tabular}} \\ \midrule
\multicolumn{2}{|c|}{\textbf{\begin{tabular}[c]{@{}c@{}}iperf (TCP)\\ \\ Training stop: 30200\\ Max validation score: 0.014\end{tabular}}} &
  \multicolumn{2}{l|}{\cellcolor[HTML]{9AFF99}\begin{tabular}[c]{@{}l@{}}Total: 199999\\      TP: 199999\\      TN: 0\\      FP: 0\\      FN: 0\\      Max Error: 2.28e19\\      Min Error: 2.29e19\end{tabular}} &
  \multicolumn{2}{l|}{\cellcolor[HTML]{9AFF99}\begin{tabular}[c]{@{}l@{}}Total: 485192\\      TP: 485192\\      TN: 0\\      FP: 0\\      FN: 0\\      Max Error: 4.18e16\\      Min Error: 4.18e16\end{tabular}} &
  \multicolumn{2}{l|}{\cellcolor[HTML]{9AFF99}\begin{tabular}[c]{@{}l@{}}Total: 80796\\      TP: 80796\\      TN: 0\\      FP: 0\\      FN: 0\\      Max Error:  2.11e19\\      Min Error: 2.11e19\end{tabular}} &
  \multicolumn{2}{l|}{\cellcolor[HTML]{9AFF99}\begin{tabular}[c]{@{}l@{}}Total: 26107\\      TP: 26107\\      TN: 0\\      FP: 0\\      FN: 0\\      Max Error:  2.12e19\\      Min Error: 2.11e19\end{tabular}} &
  \multicolumn{2}{l|}{\cellcolor[HTML]{9AFF99}\begin{tabular}[c]{@{}l@{}}Total: 134195\\      TP: 134195\\      TN: 0\\      FP: 0\\      FN: 0\\      Max Error: 2.29e19\\      Min Error: 2.28e19\end{tabular}} &
  \multicolumn{2}{l|}{\cellcolor[HTML]{9AFF99}\begin{tabular}[c]{@{}l@{}}Total: 22984\\      TP: 22984\\      TN: 0\\      FP: 0\\      FN: 0\\      Max Error: 2.29e19  \\      Min Error: 2.28e19\end{tabular}} \\ \bottomrule
\end{tabular}%
}
\end{table}

\begin{table}[]
\caption{ARL evaluation results for TCP-based services.}
\label{tab:tcp-arl-results-2}
\resizebox{\textwidth}{!}{%
\begin{tabular}{|cl|ll|l|ll|ll|ll|}
\hline
\multicolumn{2}{|c|}{\textbf{\begin{tabular}[c]{@{}c@{}}Testing (Unseen) $\rightarrow$\\ \\ Training $\downarrow$\end{tabular}}} &
  \multicolumn{2}{c|}{\textbf{2024 HTTPS 1 (TCP)}} &
  \multicolumn{1}{c|}{\textbf{2024 HTTPS 2 (TCP)}} &
  \multicolumn{2}{c|}{\textbf{iperf (TCP)}} &
  \multicolumn{2}{c|}{\textbf{SYN Flooding (TCP)}} &
  \multicolumn{2}{c|}{\textbf{PortScan (TCP)}} \\ \hline
\multicolumn{2}{|c|}{\textbf{\begin{tabular}[c]{@{}c@{}}2018 HTTP 1 (TCP)\\ \\ Training stop: 30200\\ Max validation score: 0.03\end{tabular}}} &
  \multicolumn{2}{l|}{\cellcolor[HTML]{9AFF99}\begin{tabular}[c]{@{}l@{}}Total: 866065\\      TP: 866065\\      TN: 0\\      FP: 0\\      FN: 0\\      Max Error: 1.54e18  \\      Min Error: 1.54e18\end{tabular}} &
  \cellcolor[HTML]{9AFF99}\begin{tabular}[c]{@{}l@{}}Total: 256112\\      TP: 256112\\      TN: 0\\      FP: 0\\      FN: 0\\      Max Error: 1.54e18  \\      Min Error: 1.54e18\end{tabular} &
  \multicolumn{2}{l|}{\cellcolor[HTML]{9AFF99}\begin{tabular}[c]{@{}l@{}}Total: 163917\\      TP: 163917\\      TN: 0\\      FP: 0\\      FN: 0\\      Max Error:   2.09e19\\      Min Error: 2.09e19\end{tabular}} &
  \multicolumn{2}{l|}{\cellcolor[HTML]{FFCCC9}\begin{tabular}[c]{@{}l@{}}Total: 375933\\      TP: 375933\\      TN: 0\\      FP: 0\\      FN: 0\\      Max Error:   2.09e19\\      Min Error: 2.09e19\end{tabular}} &
  \multicolumn{2}{l|}{\cellcolor[HTML]{FFCCC9}\begin{tabular}[c]{@{}l@{}}Total: 1993\\      TP: 1993\\      TN: 0\\      FP: 0\\      FN: 0\\      Max Error:   2.51e19\\      Min Error: 2.09e19\end{tabular}} \\ \hline
\multicolumn{2}{|c|}{\textbf{\begin{tabular}[c]{@{}c@{}}2018 HTTPS 1 (TCP)\\ \\ Training stop: 30200\\ Max validation score: 0.13\end{tabular}}} &
  \multicolumn{2}{l|}{\cellcolor[HTML]{9AFF99}\begin{tabular}[c]{@{}l@{}}Total: 866065\\      TP: 866065\\      TN: 0\\      FP: 0\\      FN: 0\\      Max Error: 1.44e20  \\      Min Error: 1.44e20\end{tabular}} &
  \cellcolor[HTML]{9AFF99}\begin{tabular}[c]{@{}l@{}}Total: 256112\\      TP: 256112\\      TN: 0\\      FP: 0\\      FN: 0\\      Max Error: 2.21e20  \\      Min Error: 2.21e20\end{tabular} &
  \multicolumn{2}{l|}{\cellcolor[HTML]{9AFF99}\begin{tabular}[c]{@{}l@{}}Total: 163917\\      TP: 163917\\      TN: 0\\      FP: 0\\      FN: 0\\      Max Error:   2.82e19\\      Min Error: 2.81e19\end{tabular}} &
  \multicolumn{2}{l|}{\cellcolor[HTML]{FFCCC9}\begin{tabular}[c]{@{}l@{}}Total: 375933\\      TP: 375933\\      TN: 0\\      FP: 0\\      FN: 0\\      Max Error:   1.93e19\\      Min Error: 2.73e20\end{tabular}} &
  \multicolumn{2}{l|}{\cellcolor[HTML]{FFCCC9}\begin{tabular}[c]{@{}l@{}}Total: 1993\\      TP: 1993\\      TN: 0\\      FP: 0\\      FN: 0\\      Max Error:   2.62e20\\      Min Error: 2.62e20\end{tabular}} \\ \hline
\multicolumn{2}{|c|}{\textbf{\begin{tabular}[c]{@{}c@{}}2024 HTTP 1 (TCP)\\ \\ Training stop: 30200\\ Max validation score: 0.15\end{tabular}}} &
  \multicolumn{2}{l|}{\cellcolor[HTML]{9AFF99}\begin{tabular}[c]{@{}l@{}}Total: 866065\\      TP: 866065\\      TN: 0\\      FP: 0\\      FN: 0\\      Max Error: 1.79e18  \\      Min Error: 1.79e18\end{tabular}} &
  \cellcolor[HTML]{9AFF99}\begin{tabular}[c]{@{}l@{}}Total: 256112\\      TP: 256112\\      TN: 0\\      FP: 0\\      FN: 0\\      Max Error: 1.79e18\\      Min Error: 1.79e18\end{tabular} &
  \multicolumn{2}{l|}{\cellcolor[HTML]{9AFF99}\begin{tabular}[c]{@{}l@{}}Total: 163917\\      TP: 163917\\      TN: 0\\      FP: 0\\      FN: 0\\      Max Error:  2.42e19\\      Min Error: 2.42e19\end{tabular}} &
  \multicolumn{2}{l|}{\cellcolor[HTML]{FFCCC9}\begin{tabular}[c]{@{}l@{}}Total: 375933\\      TP: 375933\\      TN: 0\\      FP: 0\\      FN: 0\\      Max Error:  2.42e19\\      Min Error:  2.42e19\end{tabular}} &
  \multicolumn{2}{l|}{\cellcolor[HTML]{FFCCC9}\begin{tabular}[c]{@{}l@{}}Total: 1993\\      TP: 1993\\      TN: 0\\      FP: 0\\      FN: 0\\      Max Error:   2.92e19\\      Min Error: 2.42e19\end{tabular}} \\ \hline
\multicolumn{2}{|c|}{\textbf{\begin{tabular}[c]{@{}c@{}}2024 HTTPS 2 (TCP)\\ \\ Training stop: 100200\\ Max validation score: 0.011\end{tabular}}} &
  \multicolumn{2}{l|}{\cellcolor[HTML]{9AFF99}\begin{tabular}[c]{@{}l@{}}Total: 866065\\      TP: 866065\\      TN: 0\\      FP: 0\\      FN: 0\\      Max Error:  4.3e15\\      Min Error: 1015\end{tabular}} &
  \cellcolor[HTML]{FFCE93}\begin{tabular}[c]{@{}l@{}}Total: 70000\\      TP: 0\\      TN: 70000\\      FP: 0\\      FN: 0\\      Max Error: 0.22  \\      Min Error: 0.003\end{tabular} &
  \multicolumn{2}{l|}{\cellcolor[HTML]{9AFF99}\begin{tabular}[c]{@{}l@{}}Total: 163917\\      TP: 163917\\      TN: 0\\      FP: 0\\      FN: 0\\      Max Error:  2e19\\      Min Error: 2e19\end{tabular}} &
  \multicolumn{2}{l|}{\cellcolor[HTML]{FFCCC9}\begin{tabular}[c]{@{}l@{}}Total: 375933\\      TP: 375933\\      TN: 0\\      FP: 0\\      FN: 0\\      Max Error:  2e19\\      Min Error:  2e19\end{tabular}} &
  \multicolumn{2}{l|}{\cellcolor[HTML]{FFCCC9}\begin{tabular}[c]{@{}l@{}}Total: 1993\\      TP: 1993\\      TN: 0\\      FP: 0\\      FN: 0\\      Max Error:   2.44e19\\      Min Error: 2e19\end{tabular}} \\ \hline
\multicolumn{2}{|c|}{\textbf{\begin{tabular}[c]{@{}c@{}}iperf (TCP)\\ \\ Training stop: 30200\\ Max validation score: 0.014\end{tabular}}} &
  \multicolumn{2}{l|}{\cellcolor[HTML]{9AFF99}\begin{tabular}[c]{@{}l@{}}Total: 866065\\      TP: 866065\\      TN: 0\\      FP: 0\\      FN: 0\\      Max Error:  2.12e19\\      Min Error: 2.11e19\end{tabular}} &
  \cellcolor[HTML]{9AFF99}\begin{tabular}[c]{@{}l@{}}Total: 256112\\      TP: 256112\\      TN: 0\\      FP: 0\\      FN: 0\\      Max Error: 2.11e19\\      Min Error:  2.11e19\end{tabular} &
  \multicolumn{2}{l|}{\cellcolor[HTML]{FFCE93}\begin{tabular}[c]{@{}l@{}}Total: 133717\\      TP: 0\\      TN: 133717\\      FP: 0\\      FN: 0\\      Max Error:  0.07\\      Min Error: 0.0002\end{tabular}} &
  \multicolumn{2}{l|}{\cellcolor[HTML]{FFCCC9}\begin{tabular}[c]{@{}l@{}}Total: 375933\\      TP: 375933\\      TN: 0\\      FP: 0\\      FN: 0\\      Max Error:  5.17e17\\      Min Error:  3.8e15\end{tabular}} &
  \multicolumn{2}{l|}{\cellcolor[HTML]{FFCCC9}\begin{tabular}[c]{@{}l@{}}Total: 1993\\      TP: 1993\\      TN: 0\\      FP: 0\\      FN: 0\\      Max Error:  4.63e18\\      Min Error: 3.9e15\end{tabular}} \\ \hline
\end{tabular}%
}
\end{table}

\subsection{RQ2: Comparing KitNet Performance With Other Models}

In this section, we present an evaluation that supports the selection of KitNet as the model of choice for the implementation of the ARL module.

We trained several baseline models to train and evaluate their performance on the individual edges of our CR graph model, including Isolation Forest, One-Class SVM, and Gaussian Mixture Model. All models, including KitNet, were trained on the same dataset and evaluated on the same test (unseen) data.

Table~\ref{tab:arl-candidate-model-results} summarizes the performance of these models, including KitNet, across four real traffic datasets derived from the MAWI traffic trace. The metric TN represents the number of instances correctly classified as normal benign traffic, while FP denotes the number of legitimate instances incorrectly flagged as anomalous. The results indicate that all baseline models exhibit suboptimal performance compared to KitNet. Notably, KitNet consistently achieved zero false positives, ensuring that no legitimate access requests are denied by the ARL module.

\begin{table}[]
\caption{Comparison of our model of choice for the ARL module against other possible candidate models.}
\label{tab:arl-candidate-model-results}
\resizebox{0.9\textwidth}{!}{%
\begin{tabular}{|cl|ll|l|ll|ll|}
\hline
\multicolumn{2}{|c|}{\textbf{Model}} &
  \multicolumn{2}{c|}{\textbf{\begin{tabular}[c]{@{}c@{}}2018 Duo 1 (UDP)\\ \\ Unseen Data Size: 203409\end{tabular}}} &
  \multicolumn{1}{c|}{\textbf{\begin{tabular}[c]{@{}c@{}}2024 Game  (UDP) \\ \\ Unseen Data Size: 39899\end{tabular}}} &
  \multicolumn{2}{c|}{\textbf{\begin{tabular}[c]{@{}c@{}}2018 Duo 4 (UDP)\\ \\ Unseen Data Size: 386575\end{tabular}}} &
  \multicolumn{2}{c|}{\textbf{\begin{tabular}[c]{@{}c@{}}2024 HTTP 1 (TCP)\\ \\ Unseen Data Size: 1170\end{tabular}}} \\ \hline
\multicolumn{2}{|c|}{\textbf{CR + Isolation Forest}} &
  \multicolumn{2}{l|}{\cellcolor[HTML]{FFFFFF}\begin{tabular}[c]{@{}l@{}}TN: 187694\\      FP: 15715\end{tabular}} &
  \cellcolor[HTML]{FFFFFF}\begin{tabular}[c]{@{}l@{}}TN: 39899\\ FP: 0\end{tabular} &
  \multicolumn{2}{l|}{\cellcolor[HTML]{FFFFFF}\begin{tabular}[c]{@{}l@{}}TN: 383428\\ FP: 3147\end{tabular}} &
  \multicolumn{2}{l|}{\cellcolor[HTML]{FFFFFF}\begin{tabular}[c]{@{}l@{}}TN: 1093\\ FP: 77\end{tabular}} \\ \hline
\multicolumn{2}{|c|}{\textbf{CR + One-Class SVM}} &
  \multicolumn{2}{l|}{\cellcolor[HTML]{FFFFFF}\begin{tabular}[c]{@{}l@{}}TN: 196637\\      FP: 6775\end{tabular}} &
  \cellcolor[HTML]{FFFFFF}\begin{tabular}[c]{@{}l@{}}TN: 39619\\ FP: 280\end{tabular} &
  \multicolumn{2}{l|}{\cellcolor[HTML]{FFFFFF}\begin{tabular}[c]{@{}l@{}}TN: 384843\\ FP: 1732\end{tabular}} &
  \multicolumn{2}{l|}{\cellcolor[HTML]{FFFFFF}\begin{tabular}[c]{@{}l@{}}TN: 0\\ FP: 1170\end{tabular}} \\ \hline
\multicolumn{2}{|c|}{\textbf{CR + Gaussian Mixture}} &
  \multicolumn{2}{l|}{\cellcolor[HTML]{FFFFFF}\begin{tabular}[c]{@{}l@{}}TN: 193260\\ FP: 10149\end{tabular}} &
  \cellcolor[HTML]{FFFFFF}\begin{tabular}[c]{@{}l@{}}TN: 37947\\ FP: 1952\end{tabular} &
  \multicolumn{2}{l|}{\cellcolor[HTML]{FFFFFF}\begin{tabular}[c]{@{}l@{}}TN: 367574\\ FP: 19001\end{tabular}} &
  \multicolumn{2}{l|}{\cellcolor[HTML]{FFFFFF}\begin{tabular}[c]{@{}l@{}}TN: 1111\\ FP: 59\end{tabular}} \\ \hline
\multicolumn{2}{|c|}{\textbf{CR + KitNet (Ours)}} &
  \multicolumn{2}{l|}{\cellcolor[HTML]{FFFFFF}\textbf{\begin{tabular}[c]{@{}l@{}}TN: 203409\\ FP: 0\end{tabular}}} &
  \cellcolor[HTML]{FFFFFF}\textbf{\begin{tabular}[c]{@{}l@{}}TN: 39899\\ FP: 0\end{tabular}} &
  \multicolumn{2}{l|}{\cellcolor[HTML]{FFFFFF}\textbf{\begin{tabular}[c]{@{}l@{}}TN: 386575\\ FP: 0\end{tabular}}} &
  \multicolumn{2}{l|}{\cellcolor[HTML]{FFFFFF}\textbf{\begin{tabular}[c]{@{}l@{}}TN: 1170\\ FP: 0\end{tabular}}} \\ \hline
\end{tabular}%
}
\end{table}

\subsection{RQ3: RTFSL Module Effectiveness}
\label{sec:rtfsl-effectiveness}

\noindent\textbf{Experiment Setup.} 
We used the same experimental setup as in Section~\ref{sec:eval-ac}.

\noindent\textbf{Model Training.} 
For this module, we trained the model using the following parameter settings: the minimum number of training flow statistics samples was 150, and the sampling rate was 5 seconds. We also set the maximum Euclidean distance threshold for anomaly detection to 0.8. Finally, we set the window size to 70 samples for UDP-based communications and 90 for TCP-based communications.

\begin{figure}[htbp]
\centerline{\includegraphics[width=.9\textwidth]{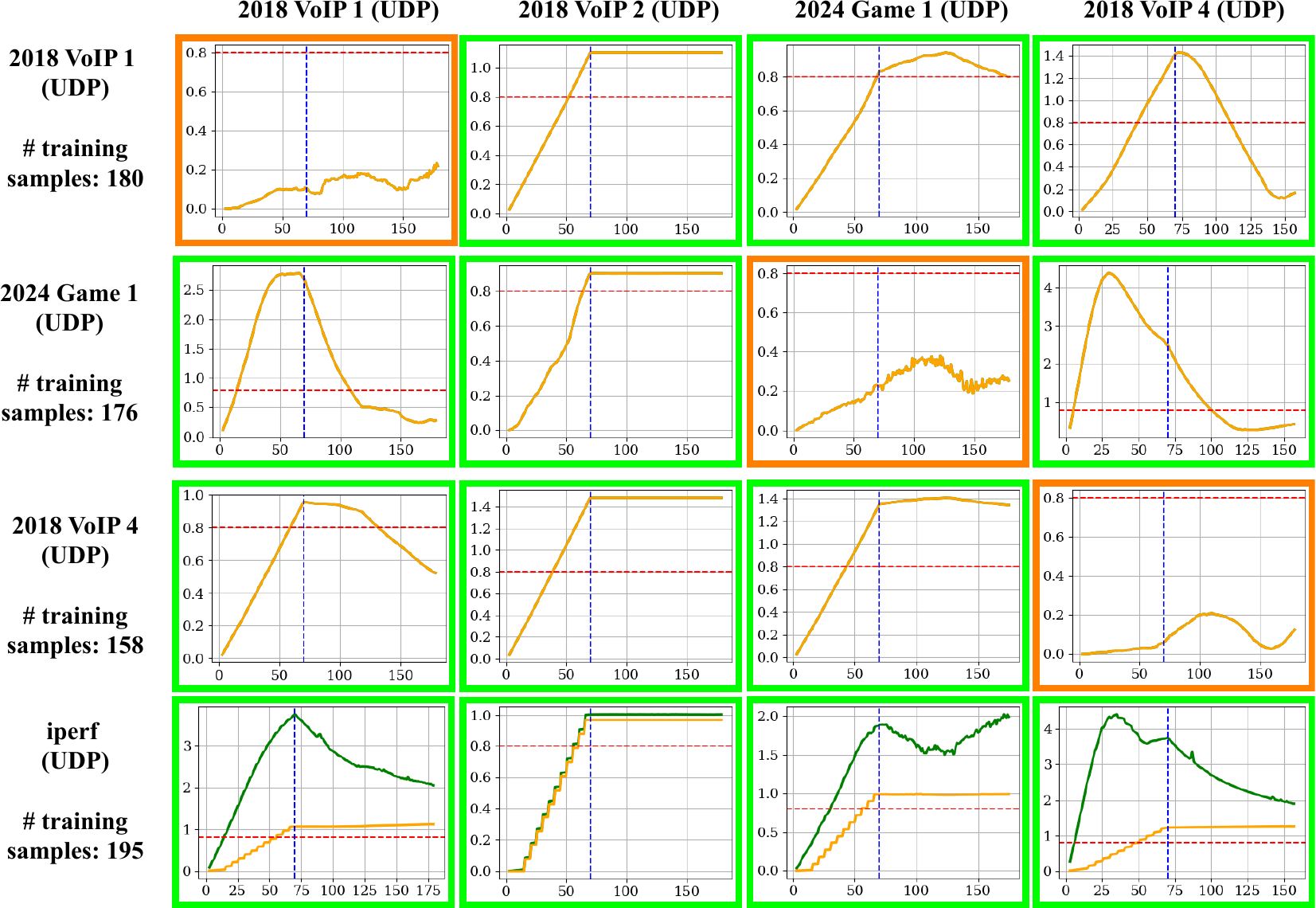}}
\caption{
RTFSL module evaluation results for the UDP-based flows.
\label{fig:rtfsl-udp-results-1}
}
\end{figure}

\noindent\textbf{Model Testing.} 
We evaluate the efficacy of the RTFSL module in ensuring that access granted to endpoints is used in accordance with the learned transmission behavior from the training process. In this experimental setup, the ARL module initially grants communication access to endpoints. However, during the enforcement phase, the endpoints deviate from their normal behavior, either adopting the transmission patterns of another benign application or exhibiting malicious behavior. 

It is important to note that the model training and enforcement phases were conducted using different traffic trace datasets captured during different periods in the MAWI dataset. These capture periods experienced natural throughput fluctuations due to the starting and stopping of flows. As a result, during testing, the model attempts to match transmission patterns that may have varied due to these throughput fluctuations. Therefore, we evaluate the performance of the RTFSL module in recognizing normality or abnormality in traffic patterns affected by the dynamic fluctuations of network conditions.

Consistent with the evaluation objectives for the ARL module, we assess the RTFSL module's ability to (1) recognize normal benign transmission patterns and (2) detect deviations classified as abnormal benign or malicious behaviors.

\begin{figure}[htbp]
\centerline{\includegraphics[width=.9\textwidth]{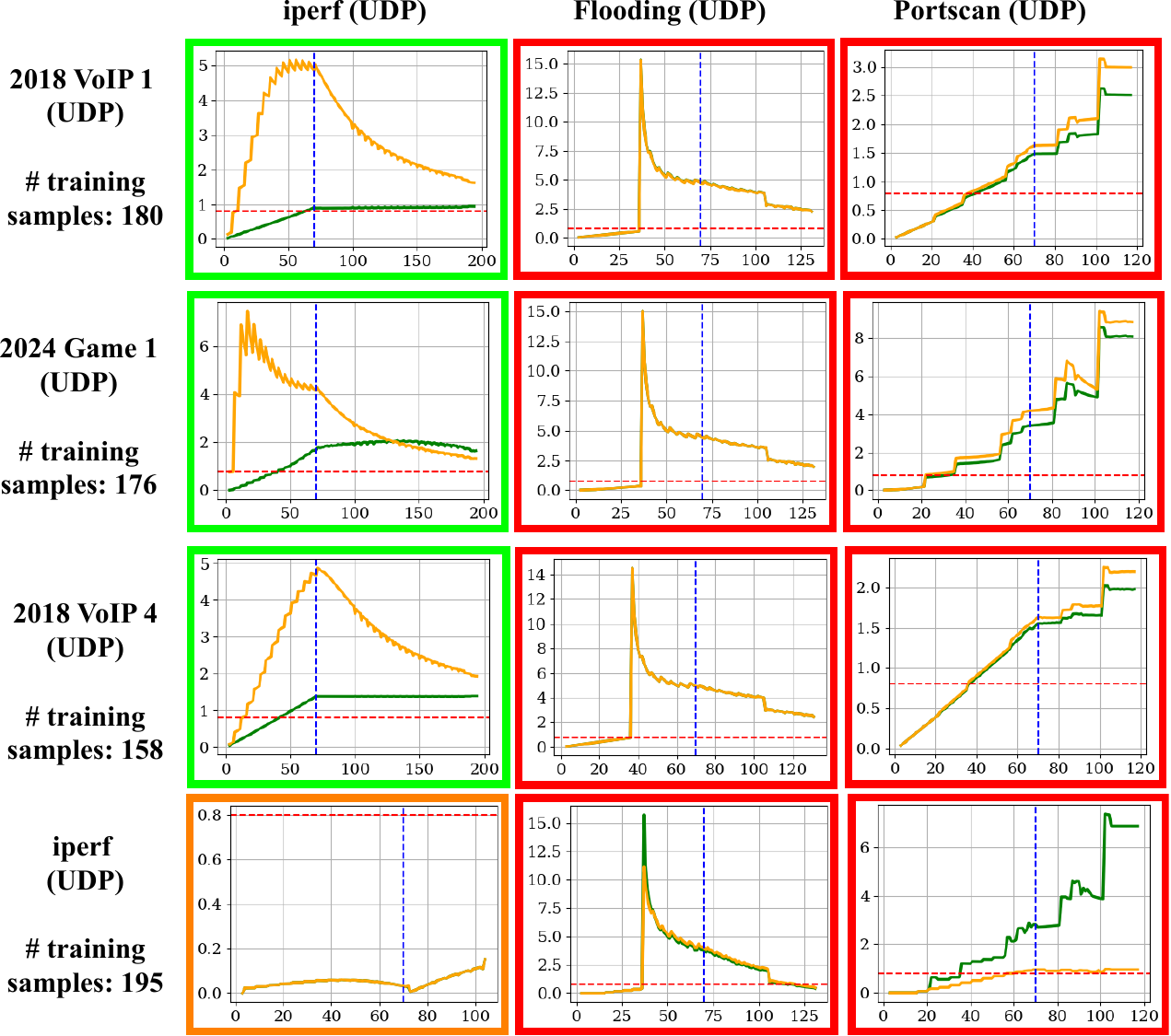}}
\caption{
RTFSL module evaluation results for the UDP-based flows (continued).
\label{fig:rtfsl-udp-results-2}
}
\end{figure}

\begin{figure}[ht]
\centerline{\includegraphics[width=1.1\textwidth]{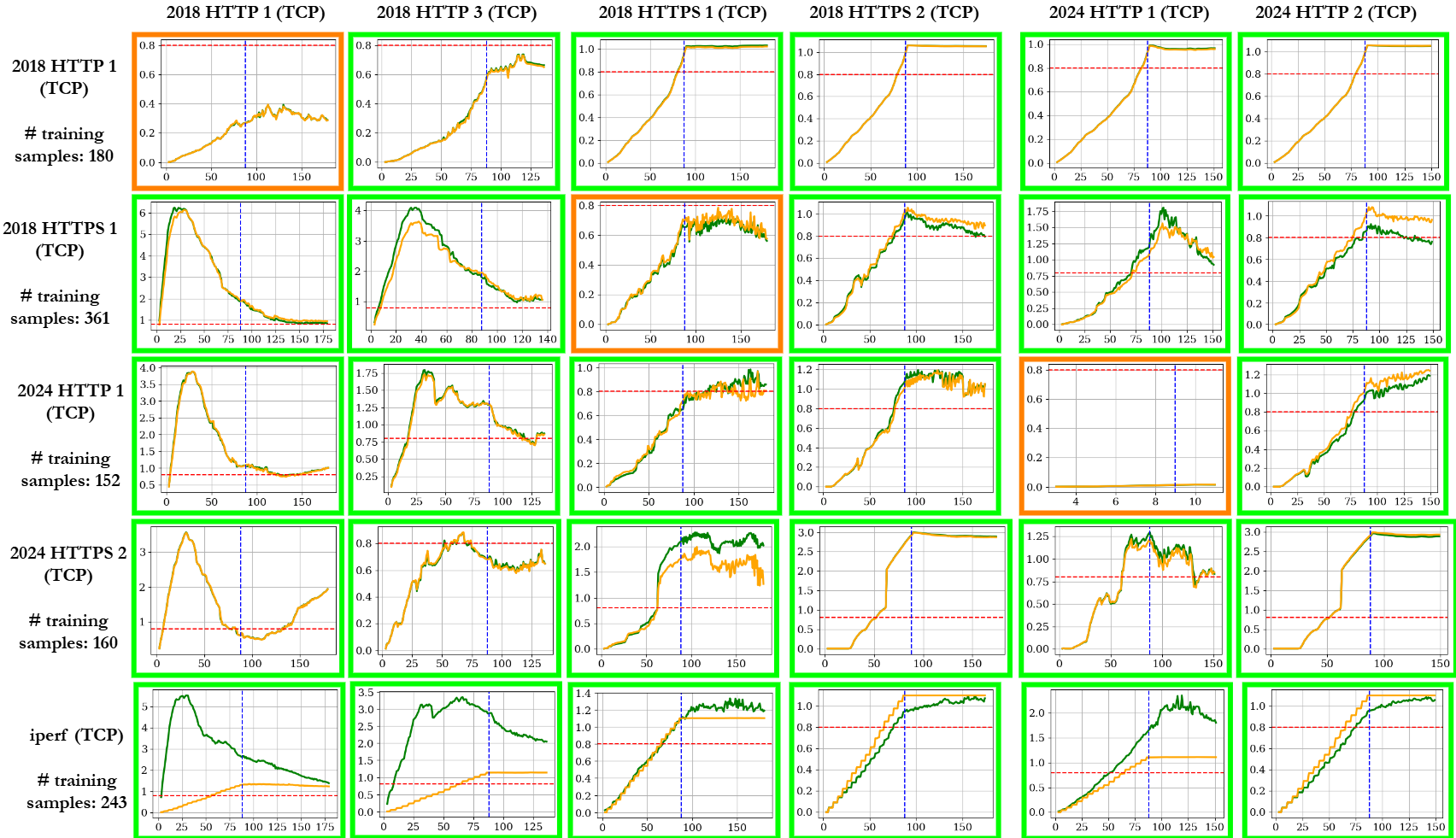}}
\caption{
RTFSL module evaluation results for the TCP-based flows.
\label{fig:rtfsl-tcp-results-1}
}
\end{figure}

\begin{figure}[htbp]
\centerline{\includegraphics[width=1.1\textwidth]{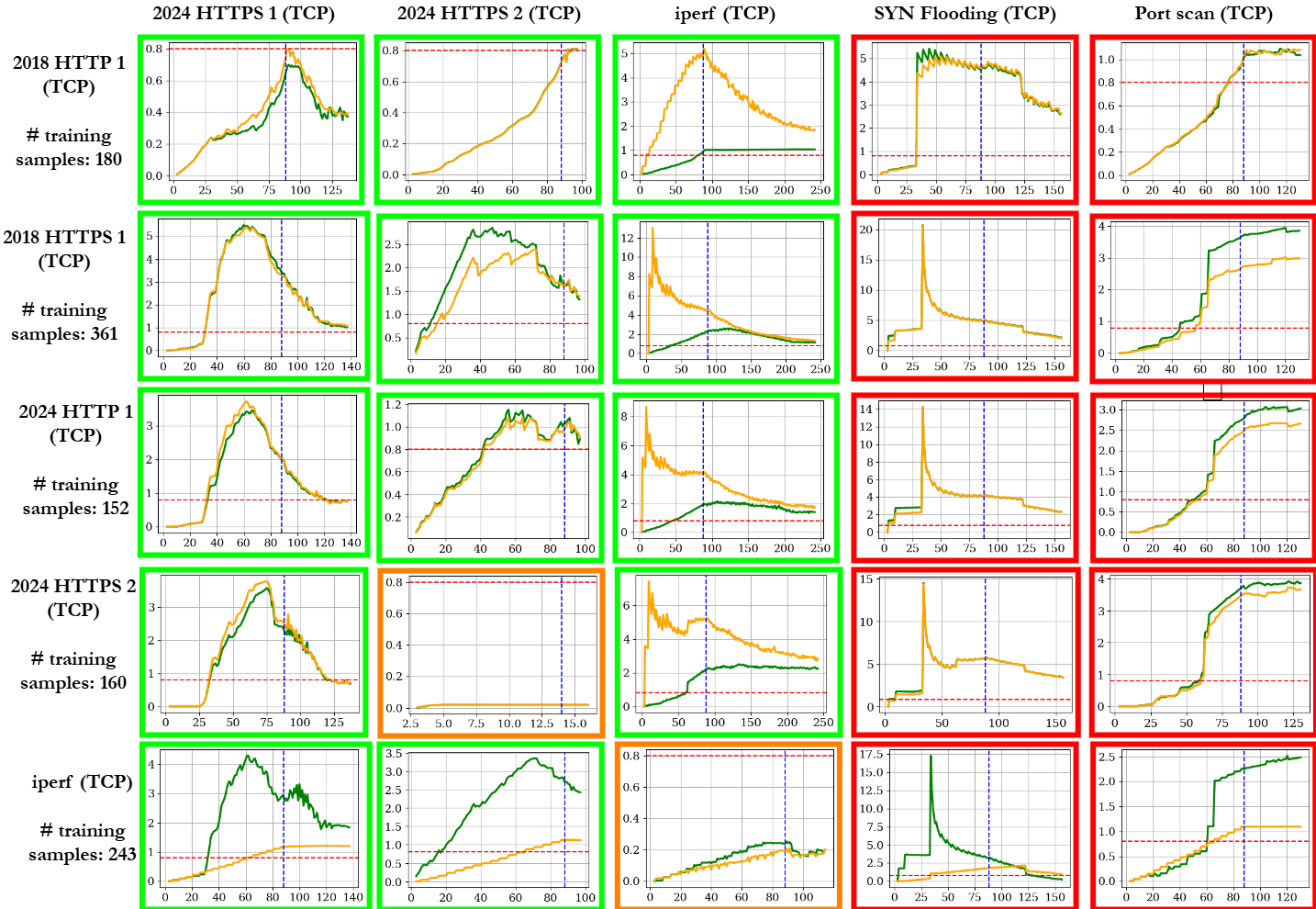}}
\caption{
RTFSL module evaluation results for the TCP-based flows (continued).
\label{fig:rtfsl-tcp-results-2}
}
\end{figure}

\noindent\textbf{Results.} Figures \ref{fig:rtfsl-udp-results-1} and \ref{fig:rtfsl-udp-results-2} show the evaluation results for the UDP-based flows whereas Figures \ref{fig:rtfsl-tcp-results-1} and \ref{fig:rtfsl-tcp-results-2} show the evaluation results for the TCP-based flows. 

Similar to the previous section, we present the results in a cross-evaluation manner where the vertical axis indicates the data on which the model is individually trained. For each training data, we report the number of training samples supplied to the model. The horizontal axis indicates the data on which the model is tested. 

Each cell in the evaluation results shows a plot of how the anomaly error (i.e., DTW Euclidean distance) progresses with the arrival of flow statistics samples for the packets (green curve) and bytes (orange curve).  The dotted blue line denotes the moment when the sliding window starts (i.e., the number of collected samples equals the window length). The horizontal dotted line denotes the anomaly detection threshold (i.e., 0.8). The anomaly detection occurs when a curve meets the anomaly detection boundary. Note that if the green and orange curves are tangent to one another, only the orange curve will be visible in the plot, as it overlays the green curve.  Plots in orange boxes pertain to normal benign tests, green to abnormal benign, and red to abnormal malicious.

In both types of flows, we make the following observations: (1) The DTW Euclidean distance for normal benign patterns typically remains far below the anomaly threshold. (2) The module detects both abnormal benign and malicious traffic patterns before the sliding window begins. In other words, in all the cases, the module detected the abnormality by collecting statistics samples less than the window length (i.e., in the first couple of minutes of data transmission). For example, the model trained on the ``2024 Game 1 (UDP)'' application data and tested on ``2018 VoIP (UDP)'' detected the anomalous flow within 10 samples (less than a minute of data transmission) from the flow start (see Figure~\ref{fig:rtfsl-udp-results-1}).

\subsection{RQ4: RTFSL Effectiveness in Changing Network Conditions}
\label{sec:rtfsl-reduced-bandwidth}

\begin{figure}[htbp]
\centerline{\includegraphics[width=\textwidth]{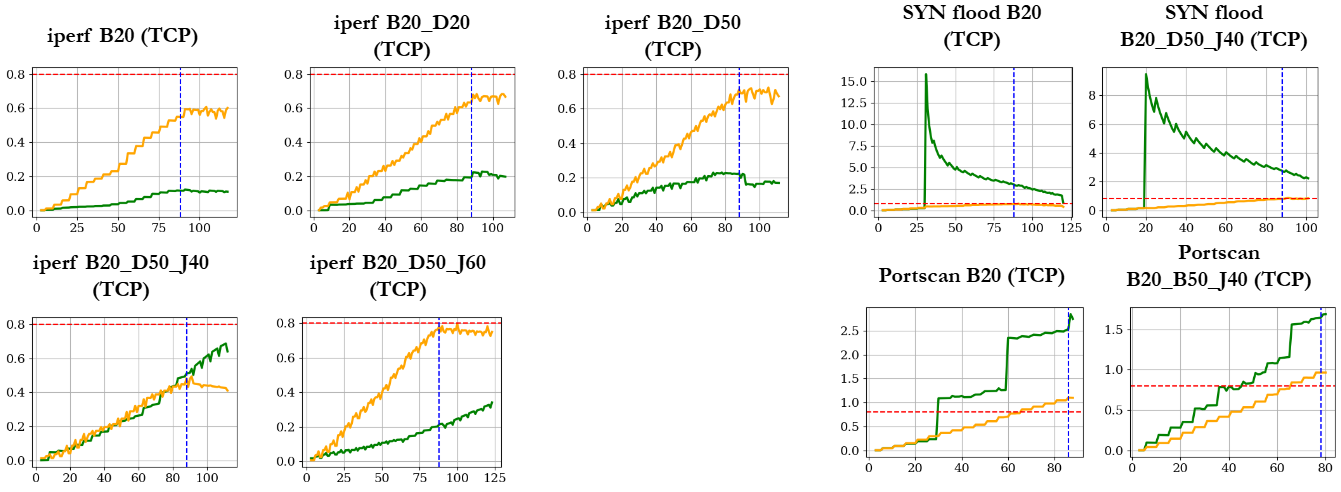}}
\caption{
RTFSL module evaluation results on different training and enforcement network conditions for the TCP-based flows.
\label{fig:rtfsl-reduced-tcp-results}
}
\end{figure}

We now explore a distinct scenario where the training data is generated under network conditions that differ from those present during model enforcement. This analysis aims to examine the module's robustness to variations in network conditions, specifically bandwidth, link delays, and jitter.

It is worth noting that the experiments conducted with the MAWI datasets (Section~\ref{sec:rtfsl-effectiveness}) inherently include fluctuations in these network condition attributes over time due to the thousands of flows traversing the transit link from which the data was captured. 
In this scenario, we intentionally introduce significant changes to these conditions to further stress-test the module's adaptability.

\noindent\textbf{Experiment Setup.} We use the experiment setup 2 discussed in Section~\ref{sec:eval-ac}. 

\begin{figure}[ht]
\centerline{\includegraphics[width=\textwidth]{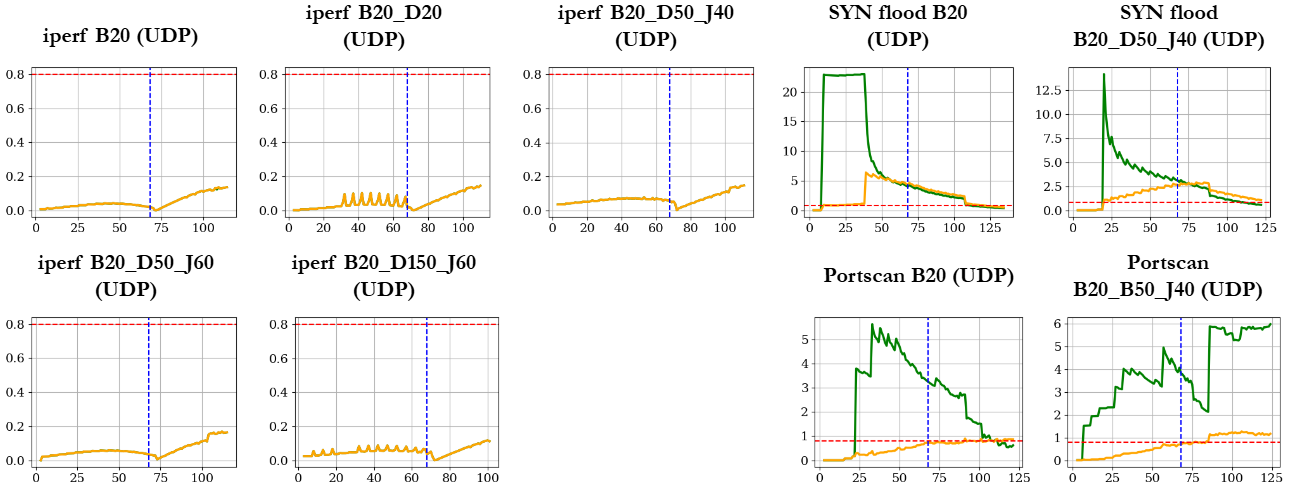}}
\caption{
RTFSL module evaluation results on different training and enforcement network conditions for the UDP-based flows.
\label{fig:rtfsl-reduced-udp-results}
}
\end{figure}

\noindent\textbf{Model Training.} 
We train the module with data generated in a network with a bandwidth of 55 Mbps between the communicating entities and no observable link delays or jitter.
In contrast, during model enforcement, we enforce the model in a network with significantly lower bandwidth (i.e., < 20 Mbps) and progressively introduce link delays and jitter. Our investigation focuses on two key aspects: (1) assessing whether normal benign transmissions are still accurately identified as benign, thereby preventing degradation of QoS by suspending benign flows; and (2) evaluating the model's ability to detect abnormal malicious flows under these changing conditions.

\noindent\textbf{Results.} The results of the experiments are presented in the Fugures \ref{fig:rtfsl-reduced-tcp-results} and \ref{fig:rtfsl-reduced-udp-results}. The name of each plot denotes the traffic patterns on which the module is tested as well as the network conditions. For instance, plot ``iperf B20\_D\_50\_J\_40,'' pertains to the exhibited transmission patterns of iperf when the bandwidth is 20 Mbps with link delays of 50 ms and jitter of 40 ms.

For both TCP- and UDP-based communication, the model demonstrates resilience against a reduction in bandwidth of at least 64\%. Specifically, the model exhibits no instances of FP or FN for either benign or anomalous malicious traffic under these conditions.

Subsequently, we introduce a network delay of 50 ms along with a jitter of up to 60 ms. In the case of TCP-based communication, the model identifies anomalies in the byte timeseries (orange curve), having an anomaly error of 0.8. Consequently, ZT-SDN terminates the connection. It is noteworthy that network delay and jitter of such magnitude significantly diminish the QoS, even without ZT-SDN intervention. Typically, applications with high bandwidth requirements or real-time communications, such as Voice over IP, are unable to tolerate jitter beyond 50 ms~\cite{ciscoMerakiJitter, mediumJitter}. Thus, the communication, even without ZT-SDN, would be impractical in such scenarios.

Conversely, in the case of UDP, the model remains robust even in the presence of extremely high delay and jitter. This resilience can be attributed to the simplicity of the UDP protocol, which lacks congestion control and retransmission mechanisms found in TCP. Consequently, the shape of the time series in these situations does not undergo notable changes that would trigger anomalies in ZT-SDN. Last, the abnormal malicious patterns exhibited by the flooding and port scanning are detectable in the reduced bandwidth setting. The model continues to exhibit robustness in detecting these attacks beyond 50 ms in both delay and jitter.

\subsection{RQ5: Scalability of ZT-SDN}
\label{sec:scalability}

We evaluate the performance of ZT-SDN by comparing it to the baseline provided by ONOS's reactive forwarding application, Fwd. Our evaluation focuses on three key metrics: (1) the number of control plane invocations (i.e., \verb|PACKET_IN| messages) from the data plane, (2) the achievable bandwidth in the data plane as a result of control plane processing, and (3) the ZT-SDN's control plane processing delay.

\textbf{Experiment Setup:} We set up a network of varying sizes in terms of hosts, switches, and the number of switches between communicating entities. Both the endpoints and switches run individually in Linux containers within the Mininet environment. We utilize \verb|Open vSwitches|~\cite{pfaff2015design} for the switches. To generate traffic and measure the maximum achievable bandwidth between each client and server machine, we use the \verb|iperf| tool in bandwidth testing mode. In this mode, \verb|iperf| uses two bidirectional channels and sends packets rapidly to stress test the connection and measure the bandwidth. In these experiments, we test the scalability of our approach under the most stressful network conditions: (1) all client machines start their applications approximately at the same time, and (2) the communication path between entities is always the longest possible, traversing all network switches.

\textbf{Model Training.} We train ARL and RTFSL similarly to Section~\ref{sec:eval-ac}. Using our RGAM approach, we mine rules from datasets generated by the CSM. The dataset comprises approximately 35,000 traffic samples from application execution. RGAM generated two flow rules for each communication endpoint: r1 = [dstAddr = $\mathit{SERVER\_ADDR}$, dstPort = 5001, dscp = 0, ecn = 0, etherType = 2048, proto = TCP, srcAddr = $\mathit{CLIENT\_ADDR}$, vlanID=-1], and r2 = [destAddr = $\mathit{CLIENT\_ADDR}$, dscp = 0, ecn = 0, etherType = 2048, proto = TCP, srcAddr = $\mathit{SERVER\_ADDR}$, srcPort = 5001, vlanID = -1], where $\mathit{SERVER\_ADDR}$ and $\mathit{CLIENT\_ADDR}$ are the IPv4 addresses the server and each client respectively.
Here, $\mathit{SERVER\_ADDR}$ and $\mathit{CLIENT\_ADDR}$ represent the IPv4 addresses of the server and each client, respectively. According to our model analysis, the source port on any client machine cannot be predicted, and thus, it is excluded from the generated rules. Additionally, the analysis indicates that these two rules have a 100\% association score.

\textbf{Results.} 
Table~\ref{tab:rgam-results}  shows the scalability performance results comparing the Fwd with ZT-SDN. The topology column reports the number of host machines (clients and server) and the number of switches in the path between each client and the server.  Packet-In indicates the number of times the controller was invoked due to the absence of rules in the data plane. Number of rules expresses the number of rules installed at the switches by each approach. Note that Fwd generates rules dynamically at runtime based on \verb|PACKET_IN| message headers, which include source and destination addresses, ports, and transport layer protocol. 
Finally, the total processing time represents the time required for ZT-SDN and Fwd to process all the access requests (i.e., \verb|PACKET_IN|). For ZT-SDN, this includes the processing time of both the CSM and ML modules (encompassing packet headers processing and model inferences). It is important to note that Fwd does not enforce access control and allows all accesses.

In all cases, ZT-SDN demonstrates a reduction in the number of \verb|PACKET_IN| messages compared to Fwd, owing to ZT-SDN's proactive rule deployment approach. Finally, the number of rules installed in the network is reduced by 50\% in ZT-SDN compared to rules installed by Fwd.
This is because Fwd enforces rules with both the source and destination ports. Since, in this scenario, there are two bidirectional channels, the number of rules enforced by Fwd at each switch would be double the number of rules enforced by ZT-SDN. 

\begin{table}[]
\centering
\caption{Number of Packet-In, rules per switch, and average processing times. \label{tab:rgam-results}}
\resizebox{.7\columnwidth}{!}{%
\begin{tabular}{c|cc|cc|cc}
\hline
\multirow{2}{*}{\textbf{\begin{tabular}[c]{@{}c@{}}Topology\\ (Hosts x Switches)\end{tabular}}} &
  \multicolumn{2}{c|}{\textbf{\begin{tabular}[c]{@{}c@{}}Total \\ Packet-In \end{tabular}}} &
  \multicolumn{2}{c|}{\textbf{\begin{tabular}[c]{@{}c@{}}Number of Rules\\ Per Switch\end{tabular}}} &
  \multicolumn{2}{c}{\textbf{\begin{tabular}[c]{@{}c@{}}Total Processing\\ Time (sec)\end{tabular}}} \\ \cline{2-7} 
                      & \textbf{ZT-SDN} & \textbf{Fwd} & \textbf{ZT-SDN} & \textbf{Fwd} & \textbf{ZT-SDN} & \textbf{Fwd} \\ \hline
\textbf{2H x 4 SW}    & 7               & 131          & 2               & 4            & 0.22            & 0.13         \\ \hline
\textbf{4H x 4 SW}    & 45              & 118          & 6               & 12           & 0.8             & 0.19         \\ \hline
\textbf{10 H x 5 SW}  & 170             & 435          & 18              & 36           & 8.86            & 0.98         \\ \hline
\textbf{20 H x 10 SW} & 1038            & 1364         & 38              & 76           & 12.22           & 4.23         \\ \hline
\textbf{30 H x 10 SW} & 1873            & 2808         & 58              & 116          & 16.87           & 5.88         \\ \hline
\textbf{40 H x 15 SW} & 2272            & 4248         & 78              & 156          & 21.78           & 8.72         \\ \hline
\end{tabular}%
}
\end{table}

Table~\ref{tab:bandwidth} shows the \verb|iperf| reported average throughput and the amount of data transmitted per host across the same network topologies reported in Table~\ref{tab:rgam-results}. 
The results suggest that ZT-SDN  does not cause network degradation despite the additional access request processing overhead at the control plane. In fact, as the network topology scales, ZT-SDN may exhibit slightly better performance than the baseline.

\begin{table}[ht]
\caption{Average maximum achievable bandwidth and bytes transmitted between the communicating entities.\label{tab:bandwidth}}
\resizebox{.7\columnwidth}{!}{%
\begin{tabular}{c|cc|cc}
\hline
\multirow{2}{*}{\textbf{\begin{tabular}[c]{@{}c@{}}Topology\\ (Hosts x Switches)\end{tabular}}} &
  \multicolumn{2}{c|}{\textbf{\begin{tabular}[c]{@{}c@{}}Average Throughput \\ Per Host (Gbps)\end{tabular}}} &
  \multicolumn{2}{c}{\textbf{\begin{tabular}[c]{@{}c@{}}Average Bytes Transmitted \\ Per Host (GBytes)\end{tabular}}} \\ \cline{2-5} 
                      & \textbf{ZT-SDN} & \textbf{Fwd} & \textbf{ZT-SDN} & \textbf{Fwd} \\ \hline
\textbf{2H x 4 SW}    & 72.8            & 70.4         & 254             & 246          \\ \hline
\textbf{4H x 4 SW}    & 27.5            & 26.7         & 95.9            & 93.1         \\ \hline
\textbf{10 H x 5 SW}  & 5.77            & 6.06         & 20.17           & 21.12        \\ \hline
\textbf{20 H x 10 SW} & 4.06            & 3.27         & 14              & 11.44        \\ \hline
\textbf{30 H x 10 SW} & 2.89            & 2.24         & 10.11           & 7.84         \\ \hline
\textbf{40 H x 15 SW} & 1.77            & 1.74         & 6.19            & 6.1          \\ \hline
\end{tabular}
}
\end{table}

%% file: conclusion.tex
\section{Conclusion}

In this paper, we have introduced ZT-SDN, a novel end-to-end ZT architecture for SDN. ZT-SDN implements automated learning processes from the underlying network that (1) capture the communication requirements of entities in the network, (2) derive the access control rules allowing the entities to execute their missions successfully in the least privilege, and (3) learn the data transmission behavior of the entities using those granted network permissions by extracting their communication patterns. Additionally, ZT-Gym is introduced to facilitate the offline generation of application datasets and training of ML modules, ensuring benign training datasets in scenarios where network data transmissions are not guaranteed to be benign.
Our experimental results demonstrate that ZT-SDN effectively prevents unauthorized access to network resources and detects deviant behaviors in entities using permitted flows. Furthermore, we show that ZT-SDN exhibits reasonable tolerance to changing network conditions. We also present the scalability performance of our framework across various topology sizes, comparing it against the ONOS baseline reactive forwarding.

%% file: main.bbl

\begin{thebibliography}{49}

%%% the Web address.

\ifx \showCODEN    \undefined \def \showCODEN     #1{\unskip}     \fi
\ifx \showDOI      \undefined \def \showDOI       #1{#1}\fi
\ifx \showISBNx    \undefined \def \showISBNx     #1{\unskip}     \fi
\ifx \showISBNxiii \undefined \def \showISBNxiii  #1{\unskip}     \fi
\ifx \showISSN     \undefined \def \showISSN      #1{\unskip}     \fi
\ifx \showLCCN     \undefined \def \showLCCN      #1{\unskip}     \fi
\ifx \shownote     \undefined \def \shownote      #1{#1}          \fi
\ifx \showarticletitle \undefined \def \showarticletitle #1{#1}   \fi
\ifx \showURL      \undefined \def \showURL       {\relax}        \fi
% invisible to TeX
\providecommand\bibfield[2]{#2}
\providecommand\bibinfo[2]{#2}
\providecommand\natexlab[1]{#1}
\providecommand\showeprint[2][]{arXiv:#2}

\bibitem[tcL({[n.\,d.]})]%
        {tcLinux}
 \bibinfo{year}{[n.\,d.]}\natexlab{}.
\newblock \bibinfo{booktitle}{\emph{tc(8) — Linux manual page}}.
\newblock
\urldef\tempurl%
\url{https://man7.org/linux/man-pages/man8/tc.8.html}
\showURL{%
\tempurl}


\bibitem[ope(2015)]%
        {openflowSwitch}
 \bibinfo{year}{2015}\natexlab{}.
\newblock \bibinfo{booktitle}{\emph{OpenFlow Switch Specification (version 1.5.1)}}.
\newblock \bibinfo{type}{{T}echnical {R}eport}. \bibinfo{institution}{Open Networking Foundation}.
\newblock


\bibitem[Abou El~Houda et~al\mbox{.}(2021)]%
        {abou2021novel}
\bibfield{author}{\bibinfo{person}{Zakaria Abou El~Houda}, \bibinfo{person}{Abdelhakim~Senhaji Hafid}, {and} \bibinfo{person}{Lyes Khoukhi}.} \bibinfo{year}{2021}\natexlab{}.
\newblock \showarticletitle{A novel machine learning framework for advanced attack detection using sdn}. In \bibinfo{booktitle}{\emph{2021 IEEE Global Communications Conference (GLOBECOM)}}. IEEE, \bibinfo{pages}{1--6}.
\newblock


\bibitem[Agarwal et~al\mbox{.}(1994)]%
        {agarwal1994fast}
\bibfield{author}{\bibinfo{person}{Rakesh Agarwal}, \bibinfo{person}{Ramakrishnan Srikant}, {et~al\mbox{.}}} \bibinfo{year}{1994}\natexlab{}.
\newblock \showarticletitle{Fast algorithms for mining association rules}. In \bibinfo{booktitle}{\emph{Proc. of the 20th VLDB Conference}}, Vol.~\bibinfo{volume}{487}. \bibinfo{pages}{499}.
\newblock


\bibitem[Alexey~Kuznetsov({[n.\,d.]})]%
        {ss}
\bibfield{author}{\bibinfo{person}{Michael~Prokop Alexey~Kuznetsov}.} \bibinfo{year}{[n.\,d.]}\natexlab{}.
\newblock \showarticletitle{ss(8) — Linux manual page}.
\newblock \bibinfo{journal}{\emph{Linux Documentation}} (\bibinfo{year}{[n.\,d.]}).
\newblock


\bibitem[Anjum et~al\mbox{.}(2022)]%
        {anjum2022removing}
\bibfield{author}{\bibinfo{person}{Iffat Anjum}, \bibinfo{person}{Daniel Kostecki}, \bibinfo{person}{Ethan Leba}, \bibinfo{person}{Jessica Sokal}, \bibinfo{person}{Rajit Bharambe}, \bibinfo{person}{William Enck}, \bibinfo{person}{Cristina Nita-Rotaru}, {and} \bibinfo{person}{Bradley Reaves}.} \bibinfo{year}{2022}\natexlab{}.
\newblock \showarticletitle{Removing the reliance on perimeters for security using network views}. In \bibinfo{booktitle}{\emph{Proceedings of the 27th ACM on Symposium on Access Control Models and Technologies}}. \bibinfo{pages}{151--162}.
\newblock


\bibitem[Anjum et~al\mbox{.}(2023)]%
        {anjum2023msnetviews}
\bibfield{author}{\bibinfo{person}{Iffat Anjum}, \bibinfo{person}{Jessica Sokal}, \bibinfo{person}{Hafiza~Ramzah Rehman}, \bibinfo{person}{Ben Weintraub}, \bibinfo{person}{Ethan Leba}, \bibinfo{person}{William Enck}, \bibinfo{person}{Cristina Nita-Rotaru}, {and} \bibinfo{person}{Bradley Reaves}.} \bibinfo{year}{2023}\natexlab{}.
\newblock \showarticletitle{MSNetViews: Geographically Distributed Management of Enterprise Network Security Policy}. In \bibinfo{booktitle}{\emph{Proceedings of the 28th ACM Symposium on Access Control Models and Technologies}}. \bibinfo{pages}{121--132}.
\newblock


\bibitem[Apiletti et~al\mbox{.}(2009)]%
        {apiletti2009characterizing}
\bibfield{author}{\bibinfo{person}{Daniele Apiletti}, \bibinfo{person}{Elena Baralis}, \bibinfo{person}{Tania Cerquitelli}, {and} \bibinfo{person}{Vincenzo D’Elia}.} \bibinfo{year}{2009}\natexlab{}.
\newblock \showarticletitle{Characterizing network traffic by means of the NetMine framework}.
\newblock \bibinfo{journal}{\emph{Computer Networks}} \bibinfo{volume}{53}, \bibinfo{number}{6} (\bibinfo{year}{2009}), \bibinfo{pages}{774--789}.
\newblock


\bibitem[Carvalho et~al\mbox{.}(2018)]%
        {carvalho2018ecosystem}
\bibfield{author}{\bibinfo{person}{Luiz~Fernando Carvalho}, \bibinfo{person}{Taufik Abr{\~a}o}, \bibinfo{person}{Leonardo de Souza~Mendes}, {and} \bibinfo{person}{Mario~Lemes Proen{\c{c}}a~Jr}.} \bibinfo{year}{2018}\natexlab{}.
\newblock \showarticletitle{An ecosystem for anomaly detection and mitigation in software-defined networking}.
\newblock \bibinfo{journal}{\emph{Expert Systems with Applications}}  \bibinfo{volume}{104} (\bibinfo{year}{2018}), \bibinfo{pages}{121--133}.
\newblock


\bibitem[Casado et~al\mbox{.}(2007)]%
        {casado2007ethane}
\bibfield{author}{\bibinfo{person}{Martin Casado}, \bibinfo{person}{Michael~J Freedman}, \bibinfo{person}{Justin Pettit}, \bibinfo{person}{Jianying Luo}, \bibinfo{person}{Nick McKeown}, {and} \bibinfo{person}{Scott Shenker}.} \bibinfo{year}{2007}\natexlab{}.
\newblock \showarticletitle{Ethane: Taking control of the enterprise}.
\newblock \bibinfo{journal}{\emph{ACM SIGCOMM computer communication review}} \bibinfo{volume}{37}, \bibinfo{number}{4} (\bibinfo{year}{2007}), \bibinfo{pages}{1--12}.
\newblock


\bibitem[Cisco({[n.\,d.]})]%
        {ciscoPortMirroring}
\bibfield{author}{\bibinfo{person}{Cisco}.} \bibinfo{year}{[n.\,d.]}\natexlab{}.
\newblock \bibinfo{booktitle}{\emph{Configuring Traffic Mirroring}}.
\newblock
\urldef\tempurl%
\url{https://www.cisco.com/c/en/us/td/docs/iosxr/ncs5000/interfaces/711x/configuration/guide/b-interfaces-hardware-component-cg-ncs5000-711x/configuring-traffic-mirroring.pdf}
\showURL{%
\tempurl}


\bibitem[Cisco(2020)]%
        {ciscoMerakiJitter}
\bibfield{author}{\bibinfo{person}{Cisco}.} \bibinfo{year}{2020}\natexlab{}.
\newblock \bibinfo{booktitle}{\emph{What is Jitter?}}
\newblock
\urldef\tempurl%
\url{https://documentation.meraki.com/MR/Wi-Fi_Basics_and_Best_Practices/What_is_Jitter%3F}
\showURL{%
\tempurl}


\bibitem[Csikor et~al\mbox{.}(2022)]%
        {csikor2022zerodns}
\bibfield{author}{\bibinfo{person}{Levente Csikor}, \bibinfo{person}{Sriram Ramachandran}, {and} \bibinfo{person}{Anantharaman Lakshminarayanan}.} \bibinfo{year}{2022}\natexlab{}.
\newblock \showarticletitle{ZeroDNS: Towards Better Zero Trust Security using DNS}. In \bibinfo{booktitle}{\emph{Proceedings of the 38th Annual Computer Security Applications Conference}}. \bibinfo{pages}{699--713}.
\newblock


\bibitem[da~Silva et~al\mbox{.}(2016)]%
        {da2016atlantic}
\bibfield{author}{\bibinfo{person}{Anderson~Santos da Silva}, \bibinfo{person}{Juliano~Araujo Wickboldt}, \bibinfo{person}{Lisandro~Zambenedetti Granville}, {and} \bibinfo{person}{Alberto Schaeffer-Filho}.} \bibinfo{year}{2016}\natexlab{}.
\newblock \showarticletitle{ATLANTIC: A framework for anomaly traffic detection, classification, and mitigation in SDN}. In \bibinfo{booktitle}{\emph{NOMS 2016-2016 IEEE/IFIP Network Operations and Management Symposium}}. IEEE, \bibinfo{pages}{27--35}.
\newblock


\bibitem[El~Sayed et~al\mbox{.}(2022)]%
        {el2022flow}
\bibfield{author}{\bibinfo{person}{Mahmoud~Said El~Sayed}, \bibinfo{person}{Nhien-An Le-Khac}, \bibinfo{person}{Marianne~A Azer}, {and} \bibinfo{person}{Anca~D Jurcut}.} \bibinfo{year}{2022}\natexlab{}.
\newblock \showarticletitle{A flow-based anomaly detection approach with feature selection method against ddos attacks in sdns}.
\newblock \bibinfo{journal}{\emph{IEEE Transactions on Cognitive Communications and Networking}} \bibinfo{volume}{8}, \bibinfo{number}{4} (\bibinfo{year}{2022}), \bibinfo{pages}{1862--1880}.
\newblock


\bibitem[Emmerich et~al\mbox{.}(2017)]%
        {emmerich2017flowscope}
\bibfield{author}{\bibinfo{person}{Paul Emmerich}, \bibinfo{person}{Maximilian Pudelko}, \bibinfo{person}{Sebastian Gallenm{\"u}ller}, {and} \bibinfo{person}{Georg Carle}.} \bibinfo{year}{2017}\natexlab{}.
\newblock \showarticletitle{FlowScope: Efficient packet capture and storage in 100 Gbit/s networks}. In \bibinfo{booktitle}{\emph{2017 IFIP Networking Conference (IFIP Networking) and Workshops}}. IEEE, \bibinfo{pages}{1--9}.
\newblock


\bibitem[Garg et~al\mbox{.}(2019)]%
        {garg2019hybrid}
\bibfield{author}{\bibinfo{person}{Sahil Garg}, \bibinfo{person}{Kuljeet Kaur}, \bibinfo{person}{Neeraj Kumar}, {and} \bibinfo{person}{Joel~JPC Rodrigues}.} \bibinfo{year}{2019}\natexlab{}.
\newblock \showarticletitle{Hybrid deep-learning-based anomaly detection scheme for suspicious flow detection in SDN: A social multimedia perspective}.
\newblock \bibinfo{journal}{\emph{IEEE Transactions on Multimedia}} \bibinfo{volume}{21}, \bibinfo{number}{3} (\bibinfo{year}{2019}), \bibinfo{pages}{566--578}.
\newblock


\bibitem[Giotis et~al\mbox{.}(2014)]%
        {giotis2014combining}
\bibfield{author}{\bibinfo{person}{Kostas Giotis}, \bibinfo{person}{Christos Argyropoulos}, \bibinfo{person}{Georgios Androulidakis}, \bibinfo{person}{Dimitrios Kalogeras}, {and} \bibinfo{person}{Vasilis Maglaris}.} \bibinfo{year}{2014}\natexlab{}.
\newblock \showarticletitle{Combining OpenFlow and sFlow for an effective and scalable anomaly detection and mitigation mechanism on SDN environments}.
\newblock \bibinfo{journal}{\emph{Computer Networks}}  \bibinfo{volume}{62} (\bibinfo{year}{2014}), \bibinfo{pages}{122--136}.
\newblock


\bibitem[Golnabi et~al\mbox{.}(2006)]%
        {golnabi2006analysis}
\bibfield{author}{\bibinfo{person}{Korosh Golnabi}, \bibinfo{person}{Richard~K Min}, \bibinfo{person}{Latifur Khan}, {and} \bibinfo{person}{Ehab Al-Shaer}.} \bibinfo{year}{2006}\natexlab{}.
\newblock \showarticletitle{Analysis of firewall policy rules using data mining techniques}. In \bibinfo{booktitle}{\emph{2006 IEEE/IFIP Network Operations and Management Symposium NOMS 2006}}. IEEE, \bibinfo{pages}{305--315}.
\newblock


\bibitem[Holland et~al\mbox{.}(2021)]%
        {holland2021new}
\bibfield{author}{\bibinfo{person}{Jordan Holland}, \bibinfo{person}{Paul Schmitt}, \bibinfo{person}{Nick Feamster}, {and} \bibinfo{person}{Prateek Mittal}.} \bibinfo{year}{2021}\natexlab{}.
\newblock \showarticletitle{New directions in automated traffic analysis}. In \bibinfo{booktitle}{\emph{Proceedings of the 2021 ACM SIGSAC Conference on Computer and Communications Security}}. \bibinfo{pages}{3366--3383}.
\newblock


\bibitem[Isyaku et~al\mbox{.}(2020)]%
        {isyaku2020software}
\bibfield{author}{\bibinfo{person}{Babangida Isyaku}, \bibinfo{person}{Mohd~Soperi Mohd~Zahid}, \bibinfo{person}{Maznah Bte~Kamat}, \bibinfo{person}{Kamalrulnizam Abu~Bakar}, {and} \bibinfo{person}{Fuad~A Ghaleb}.} \bibinfo{year}{2020}\natexlab{}.
\newblock \showarticletitle{Software defined networking flow table management of openflow switches performance and security challenges: A survey}.
\newblock \bibinfo{journal}{\emph{Future Internet}} \bibinfo{volume}{12}, \bibinfo{number}{9} (\bibinfo{year}{2020}), \bibinfo{pages}{147}.
\newblock


\bibitem[Jian-Ping et~al\mbox{.}(2012)]%
        {robotf}
\bibfield{author}{\bibinfo{person}{Liu Jian-Ping}, \bibinfo{person}{Liu Juan-Juan}, {and} \bibinfo{person}{Wang Dong-Long}.} \bibinfo{year}{2012}\natexlab{}.
\newblock \showarticletitle{Application Analysis of Automated Testing Framework Based on Robot}.
\newblock  (\bibinfo{year}{2012}), \bibinfo{pages}{194--197}.
\newblock
\urldef\tempurl%
\url{https://doi.org/10.1109/ICNDC.2012.53}
\showDOI{\tempurl}


\bibitem[Katsis et~al\mbox{.}(2021)]%
        {katsis2021can}
\bibfield{author}{\bibinfo{person}{Charalampos Katsis}, \bibinfo{person}{Fabrizio Cicala}, \bibinfo{person}{Dan Thomsen}, \bibinfo{person}{Nathan Ringo}, {and} \bibinfo{person}{Elisa Bertino}.} \bibinfo{year}{2021}\natexlab{}.
\newblock \showarticletitle{Can i reach you? do i need to? new semantics in security policy specification and testing}. In \bibinfo{booktitle}{\emph{Proceedings of the 26th ACM Symposium on Access Control Models and Technologies}}. \bibinfo{pages}{165--174}.
\newblock


\bibitem[Katsis et~al\mbox{.}(2022)]%
        {katsis2022neutron}
\bibfield{author}{\bibinfo{person}{Charalampos Katsis}, \bibinfo{person}{Fabrizio Cicala}, \bibinfo{person}{Dan Thomsen}, \bibinfo{person}{Nathan Ringo}, {and} \bibinfo{person}{Elisa Bertino}.} \bibinfo{year}{2022}\natexlab{}.
\newblock \showarticletitle{Neutron: A graph-based pipeline for zero-trust network architectures}. In \bibinfo{booktitle}{\emph{Proceedings of the Twelfth ACM Conference on Data and Application Security and Privacy}}. \bibinfo{pages}{167--178}.
\newblock


\bibitem[Kreutz et~al\mbox{.}(2014)]%
        {kreutz2014software}
\bibfield{author}{\bibinfo{person}{Diego Kreutz}, \bibinfo{person}{Fernando~MV Ramos}, \bibinfo{person}{Paulo~Esteves Verissimo}, \bibinfo{person}{Christian~Esteve Rothenberg}, \bibinfo{person}{Siamak Azodolmolky}, {and} \bibinfo{person}{Steve Uhlig}.} \bibinfo{year}{2014}\natexlab{}.
\newblock \showarticletitle{Software-defined networking: A comprehensive survey}.
\newblock \bibinfo{journal}{\emph{Proc. IEEE}} \bibinfo{volume}{103}, \bibinfo{number}{1} (\bibinfo{year}{2014}), \bibinfo{pages}{14--76}.
\newblock


\bibitem[Kumbhare and Chobe(2014)]%
        {kumbhare2014overview}
\bibfield{author}{\bibinfo{person}{Trupti~A Kumbhare} {and} \bibinfo{person}{Santosh~V Chobe}.} \bibinfo{year}{2014}\natexlab{}.
\newblock \showarticletitle{An overview of association rule mining algorithms}.
\newblock \bibinfo{journal}{\emph{International Journal of Computer Science and Information Technologies}} \bibinfo{volume}{5}, \bibinfo{number}{1} (\bibinfo{year}{2014}), \bibinfo{pages}{927--930}.
\newblock


\bibitem[Lunardi et~al\mbox{.}(2022)]%
        {lunardi2022arcade}
\bibfield{author}{\bibinfo{person}{Willian~Tessaro Lunardi}, \bibinfo{person}{Martin~Andreoni Lopez}, {and} \bibinfo{person}{Jean-Pierre Giacalone}.} \bibinfo{year}{2022}\natexlab{}.
\newblock \showarticletitle{Arcade: Adversarially regularized convolutional autoencoder for network anomaly detection}.
\newblock \bibinfo{journal}{\emph{IEEE Transactions on Network and Service Management}} \bibinfo{volume}{20}, \bibinfo{number}{2} (\bibinfo{year}{2022}), \bibinfo{pages}{1305--1318}.
\newblock


\bibitem[Mandara~Nagendra(2018)]%
        {robot-boon}
\bibfield{author}{\bibinfo{person}{Dr. T H~Sreenivas Mandara~Nagendra, C N~Chinnaswamy}.} \bibinfo{year}{November 2018}\natexlab{}.
\newblock \showarticletitle{Robot Framework: A boon for Automation}.
\newblock \bibinfo{journal}{\emph{IJSDR | Volume 3, Issue 11}} (\bibinfo{year}{November 2018}).
\newblock


\bibitem[McKeown et~al\mbox{.}(2008)]%
        {mckeown2008openflow}
\bibfield{author}{\bibinfo{person}{Nick McKeown}, \bibinfo{person}{Tom Anderson}, \bibinfo{person}{Hari Balakrishnan}, \bibinfo{person}{Guru Parulkar}, \bibinfo{person}{Larry Peterson}, \bibinfo{person}{Jennifer Rexford}, \bibinfo{person}{Scott Shenker}, {and} \bibinfo{person}{Jonathan Turner}.} \bibinfo{year}{2008}\natexlab{}.
\newblock \showarticletitle{OpenFlow: enabling innovation in campus networks}.
\newblock \bibinfo{journal}{\emph{ACM SIGCOMM computer communication review}} \bibinfo{volume}{38}, \bibinfo{number}{2} (\bibinfo{year}{2008}), \bibinfo{pages}{69--74}.
\newblock


\bibitem[Medium(2016)]%
        {mediumJitter}
\bibfield{author}{\bibinfo{person}{Medium}.} \bibinfo{year}{2016}\natexlab{}.
\newblock \bibinfo{booktitle}{\emph{What is Acceptable Jitter?}}
\newblock
\urldef\tempurl%
\url{https://medium.com/@datapath_io/what-is-acceptable-jitter-7e93c1e68f9b}
\showURL{%
\tempurl}


\bibitem[Mirsky et~al\mbox{.}(2018)]%
        {mirsky2018kitsune}
\bibfield{author}{\bibinfo{person}{Yisroel Mirsky}, \bibinfo{person}{Tomer Doitshman}, \bibinfo{person}{Yuval Elovici}, {and} \bibinfo{person}{Asaf Shabtai}.} \bibinfo{year}{2018}\natexlab{}.
\newblock \showarticletitle{Kitsune: an ensemble of autoencoders for online network intrusion detection}.
\newblock \bibinfo{journal}{\emph{arXiv preprint arXiv:1802.09089}} (\bibinfo{year}{2018}).
\newblock


\bibitem[Nanda et~al\mbox{.}(2016)]%
        {nanda2016predicting}
\bibfield{author}{\bibinfo{person}{Saurav Nanda}, \bibinfo{person}{Faheem Zafari}, \bibinfo{person}{Casimer DeCusatis}, \bibinfo{person}{Eric Wedaa}, {and} \bibinfo{person}{Baijian Yang}.} \bibinfo{year}{2016}\natexlab{}.
\newblock \showarticletitle{Predicting network attack patterns in SDN using machine learning approach}. In \bibinfo{booktitle}{\emph{2016 IEEE Conference on Network Function Virtualization and Software Defined Networks (NFV-SDN)}}. IEEE, \bibinfo{pages}{167--172}.
\newblock


\bibitem[Nayak et~al\mbox{.}(2009)]%
        {nayak2009resonance}
\bibfield{author}{\bibinfo{person}{Ankur~Kumar Nayak}, \bibinfo{person}{Alex Reimers}, \bibinfo{person}{Nick Feamster}, {and} \bibinfo{person}{Russ Clark}.} \bibinfo{year}{2009}\natexlab{}.
\newblock \showarticletitle{Resonance: Dynamic access control for enterprise networks}. In \bibinfo{booktitle}{\emph{Proceedings of the 1st ACM workshop on Research on enterprise networking}}. \bibinfo{pages}{11--18}.
\newblock


\bibitem[Nelson et~al\mbox{.}(2010)]%
        {nelson2010margrave}
\bibfield{author}{\bibinfo{person}{Timothy Nelson}, \bibinfo{person}{Christopher Barratt}, \bibinfo{person}{Daniel~J Dougherty}, \bibinfo{person}{Kathi Fisler}, {and} \bibinfo{person}{Shriram Krishnamurthi}.} \bibinfo{year}{2010}\natexlab{}.
\newblock \showarticletitle{The margrave tool for firewall analysis}. In \bibinfo{booktitle}{\emph{Proceedings of the 24th Large Installation System Administration Conference (LISA 10)}}.
\newblock


\bibitem[Nguyen et~al\mbox{.}(2008)]%
        {nguyen2008network}
\bibfield{author}{\bibinfo{person}{Huy~Anh Nguyen}, \bibinfo{person}{Tam Van~Nguyen}, \bibinfo{person}{Dong~Il Kim}, {and} \bibinfo{person}{Deokjai Choi}.} \bibinfo{year}{2008}\natexlab{}.
\newblock \showarticletitle{Network traffic anomalies detection and identification with flow monitoring}. In \bibinfo{booktitle}{\emph{2008 5th IFIP International Conference on Wireless and Optical Communications Networks (WOCN'08)}}. IEEE, \bibinfo{pages}{1--5}.
\newblock


\bibitem[ONF({[n.\,d.]})]%
        {onosTopoService}
\bibfield{author}{\bibinfo{person}{ONF}.} \bibinfo{year}{[n.\,d.]}\natexlab{}.
\newblock \bibinfo{booktitle}{\emph{Interface TopologyService}}.
\newblock
\urldef\tempurl%
\url{https://api.onosproject.org/2.7.0/apidocs/org/onosproject/net/topology/TopologyService.html}
\showURL{%
\tempurl}


\bibitem[Peng et~al\mbox{.}(2018)]%
        {peng2018detection}
\bibfield{author}{\bibinfo{person}{Huijun Peng}, \bibinfo{person}{Zhe Sun}, \bibinfo{person}{Xuejian Zhao}, \bibinfo{person}{Shuhua Tan}, {and} \bibinfo{person}{Zhixin Sun}.} \bibinfo{year}{2018}\natexlab{}.
\newblock \showarticletitle{A detection method for anomaly flow in software defined network}.
\newblock \bibinfo{journal}{\emph{IEEE Access}}  \bibinfo{volume}{6} (\bibinfo{year}{2018}), \bibinfo{pages}{27809--27817}.
\newblock


\bibitem[Pfaff et~al\mbox{.}(2015)]%
        {pfaff2015design}
\bibfield{author}{\bibinfo{person}{Ben Pfaff}, \bibinfo{person}{Justin Pettit}, \bibinfo{person}{Teemu Koponen}, \bibinfo{person}{Ethan Jackson}, \bibinfo{person}{Andy Zhou}, \bibinfo{person}{Jarno Rajahalme}, \bibinfo{person}{Jesse Gross}, \bibinfo{person}{Alex Wang}, \bibinfo{person}{Joe Stringer}, \bibinfo{person}{Pravin Shelar}, {et~al\mbox{.}}} \bibinfo{year}{2015}\natexlab{}.
\newblock \showarticletitle{The design and implementation of open $\{$vSwitch$\}$}. In \bibinfo{booktitle}{\emph{12th USENIX symposium on networked systems design and implementation (NSDI 15)}}. \bibinfo{pages}{117--130}.
\newblock


\bibitem[Salvador and Chan(2007)]%
        {salvador2007toward}
\bibfield{author}{\bibinfo{person}{Stan Salvador} {and} \bibinfo{person}{Philip Chan}.} \bibinfo{year}{2007}\natexlab{}.
\newblock \showarticletitle{Toward accurate dynamic time warping in linear time and space}.
\newblock \bibinfo{journal}{\emph{Intelligent Data Analysis}} \bibinfo{volume}{11}, \bibinfo{number}{5} (\bibinfo{year}{2007}), \bibinfo{pages}{561--580}.
\newblock


\bibitem[Scaranti et~al\mbox{.}(2022)]%
        {scaranti2022unsupervised}
\bibfield{author}{\bibinfo{person}{Gustavo~Frigo Scaranti}, \bibinfo{person}{Luiz~Fernando Carvalho}, \bibinfo{person}{Sylvio~Barbon Junior}, \bibinfo{person}{Jaime Lloret}, {and} \bibinfo{person}{Mario~Lemes Proen{\c{c}}a~Jr}.} \bibinfo{year}{2022}\natexlab{}.
\newblock \showarticletitle{Unsupervised online anomaly detection in Software Defined Network environments}.
\newblock \bibinfo{journal}{\emph{Expert Systems with Applications}}  \bibinfo{volume}{191} (\bibinfo{year}{2022}), \bibinfo{pages}{116225}.
\newblock


\bibitem[Singh et~al\mbox{.}(2013)]%
        {singh2013improving}
\bibfield{author}{\bibinfo{person}{Jaishree Singh}, \bibinfo{person}{Hari Ram}, {and} \bibinfo{person}{Dr~JS Sodhi}.} \bibinfo{year}{2013}\natexlab{}.
\newblock \showarticletitle{Improving efficiency of apriori algorithm using transaction reduction}.
\newblock \bibinfo{journal}{\emph{International Journal of Scientific and Research Publications}} \bibinfo{volume}{3}, \bibinfo{number}{1} (\bibinfo{year}{2013}), \bibinfo{pages}{1--4}.
\newblock


\bibitem[{SpeedGuide.net}(2024)]%
        {speedguide}
\bibfield{author}{\bibinfo{person}{{SpeedGuide.net}}.} \bibinfo{year}{2024}\natexlab{}.
\newblock \bibinfo{title}{TCP/IP Ports and Protocols Database}.
\newblock
\newblock
\urldef\tempurl%
\url{https://www.speedguide.net/ports.php}
\showURL{%
\tempurl}
\newblock
\shownote{Accessed: 2024-11-15}.


\bibitem[Stafford(2020)]%
        {stafford2020zero}
\bibfield{author}{\bibinfo{person}{VA Stafford}.} \bibinfo{year}{2020}\natexlab{}.
\newblock \showarticletitle{Zero trust architecture}.
\newblock \bibinfo{journal}{\emph{NIST special publication}}  \bibinfo{volume}{800} (\bibinfo{year}{2020}), \bibinfo{pages}{207}.
\newblock


\bibitem[Tongaonkar et~al\mbox{.}(2007)]%
        {tongaonkar2007inferring}
\bibfield{author}{\bibinfo{person}{Alok Tongaonkar}, \bibinfo{person}{Niranjan Inamdar}, {and} \bibinfo{person}{R Sekar}.} \bibinfo{year}{2007}\natexlab{}.
\newblock \showarticletitle{Inferring Higher Level Policies from Firewall Rules.}. In \bibinfo{booktitle}{\emph{LISA}}, Vol.~\bibinfo{volume}{7}. \bibinfo{pages}{1--10}.
\newblock


\bibitem[Tschannen et~al\mbox{.}(2018)]%
        {tschannen2018recent}
\bibfield{author}{\bibinfo{person}{Michael Tschannen}, \bibinfo{person}{Olivier Bachem}, {and} \bibinfo{person}{Mario Lucic}.} \bibinfo{year}{2018}\natexlab{}.
\newblock \showarticletitle{Recent advances in autoencoder-based representation learning}.
\newblock \bibinfo{journal}{\emph{arXiv preprint arXiv:1812.05069}} (\bibinfo{year}{2018}).
\newblock


\bibitem[Vanickis et~al\mbox{.}(2018)]%
        {vanickis2018access}
\bibfield{author}{\bibinfo{person}{Romans Vanickis}, \bibinfo{person}{Paul Jacob}, \bibinfo{person}{Sohelia Dehghanzadeh}, {and} \bibinfo{person}{Brian Lee}.} \bibinfo{year}{2018}\natexlab{}.
\newblock \showarticletitle{Access control policy enforcement for zero-trust-networking}. In \bibinfo{booktitle}{\emph{2018 29th Irish Signals and Systems Conference (ISSC)}}. IEEE, \bibinfo{pages}{1--6}.
\newblock


\bibitem[{WIDE Project}(2023)]%
        {mawi_wide_dataset}
\bibfield{author}{\bibinfo{person}{{WIDE Project}}.} \bibinfo{year}{2023}\natexlab{}.
\newblock \bibinfo{title}{{MAWI Working Group Traffic Archive}}.
\newblock
\newblock
\urldef\tempurl%
\url{https://mawi.wide.ad.jp/mawi/}
\showURL{%
\tempurl}
\newblock
\shownote{Accessed: 2024-10-24}.


\bibitem[Yang et~al\mbox{.}(2004)]%
        {yang2004forwarding}
\bibfield{author}{\bibinfo{person}{Lily Yang}, \bibinfo{person}{Ram Dantu}, \bibinfo{person}{Terry Anderson}, {and} \bibinfo{person}{Ram Gopal}.} \bibinfo{year}{2004}\natexlab{}.
\newblock \bibinfo{booktitle}{\emph{Forwarding and control element separation (ForCES) framework}}.
\newblock \bibinfo{type}{{T}echnical {R}eport}.
\newblock


\bibitem[Zavrak and {\.I}skefiyeli(2020)]%
        {zavrak2020anomaly}
\bibfield{author}{\bibinfo{person}{Sultan Zavrak} {and} \bibinfo{person}{Murat {\.I}skefiyeli}.} \bibinfo{year}{2020}\natexlab{}.
\newblock \showarticletitle{Anomaly-based intrusion detection from network flow features using variational autoencoder}.
\newblock \bibinfo{journal}{\emph{IEEE Access}}  \bibinfo{volume}{8} (\bibinfo{year}{2020}), \bibinfo{pages}{108346--108358}.
\newblock


\end{thebibliography}
